\documentclass[11pt,a4paper]{article}
\usepackage{silence} % to silence warnings
\usepackage{jheppub}
\usepackage[english]{babel}
\usepackage[utf8]{inputenc}
\usepackage{graphicx}
\usepackage{orcidlink} % ORCID place after xcolor pkg
\usepackage{amsmath,amssymb,amsfonts,amsthm}
\usepackage{bm,bbm}
\usepackage{tikz,stackrel}
\usepackage{pstricks}
\usepackage{pifont} %for ding symbols
\usepackage{empheq}
\newcommand\nn{\nonumber\\}

\newcommand{\bc}{\begin{center}}
\newcommand{\ec}{\end{center}}
\newcommand{\be}{\begin{equation}}
\newcommand{\ee}{\end{equation}}
\newcommand{\ba}{\begin{eqnarray}}
\newcommand{\ea}{\end{eqnarray}}
\def\bs{\begin{subequations}}
\def\es{\end{subequations}}
\newcommand{\ben}{\begin{equation*}}
\newcommand{\een}{\end{equation*}}
\newcommand{\ban}{\begin{eqnarray*}}
\newcommand{\ean}{\end{eqnarray*}}
\renewcommand{\leq}{\leqslant}
\renewcommand{\geq}{\geqslant}

\def\de{\delta}
\def\g{\gamma}
\def\la{\lambda}

\def\e{\epsilon}
\def\ve{\varepsilon}
\def\Om{\Omega}
\def\om{\omega}

\def\G{\Gamma}
\def\t{\theta}  
\def\s{\sigma}

\def\vp{\varphi}

\def\cB{\mathcal{B}}
\def\cC{\mathcal{C}}

\def\cI{\mathcal{I}}

\def\cL{\mathcal{L}}

\def\cS{\mathcal{S}}

\def\H{{\rm H}}

\def\p{\partial}

\def\B{\Box}

\newcommand{\Eq}[1]{(\ref{#1})}
\newcommand{\Eqq}[1]{eq.~(\ref{#1})}
\newcommand{\Eqqs}[1]{eqs.~(\ref{#1})}

\def\cob{\color{blue}}

\newcommand{\au}[2]{#1.~#2}
\newcommand{\book}[5]{\emph{#1}, #2, #3, #4 (#5)}

\newcommand{\oarX}[1]{\href{http://arxiv.org/abs/#1}{{\ttfamily\cob arXiv:#1}}}
\newcommand{\arX}[1]{\href{http://arxiv.org/abs/#1}{{\ttfamily\cob arXiv:#1}}}
\newcommand{\doin}[6]{\href{http://dx.doi.org/#1}{{\cob {\it #2} {\bf #3 #4} (#6) #5}}}
\newcommand{\doinn}[5]{\href{http://dx.doi.org/#1}{{\cob {\it #2} {\bf #3} (#5) #4}}}
\newcommand{\doij}[5]{\href{http://dx.doi.org/#1}{{\cob {\it #2} {\bf #3} (#5) #4}}}

\newcommand{\ndoinn}[5]{\href{#1}{{\cob {\it #2} {\bf #3} (#5) #4}}}
\newcommand{\procsinm}[5]{in \emph{#1}, #2 (eds.), #3, #4 (#5)}

\newcommand{\tia}[1]{\textit{#1},}
\newcommand{\boxd}[1]{\boxed{\phantom{\Biggl(}#1\phantom{\Biggl)}}}

\def\lpl{\ell_{\rm Pl}}

\def\lst{\ell_*}
\def\Lst{\Lambda_*}
\def\rme{e}
\def\rmd{d}
\def\rmi{i}
\def\Re{{\rm Re}}
\def\Im{{\rm Im}}

\newcounter{listcounter}

\allowdisplaybreaks

\begin{document}
	
\renewcommand{\thefootnote}{\fnsymbol{footnote}}

\title{Form factors, spectral and K\"all\'en--Lehmann representation in nonlocal quantum gravity}

\author[a,b,c]{Fabio Briscese\,\orcidlink{0000-0002-9519-5896},}
\emailAdd{fabio.briscese@uniroma3.it}
\affiliation[a]{Dipartimento di Architettura, Università Roma Tre, Via Aldo Manuzio 68L, 00153 Rome, Italy}
\affiliation[b]{Istituto Nazionale di Alta Matematica Francesco Severi, Gruppo
Nazionale di Fisica Matematica, Piazzale Aldo Moro 5, 00185 Rome, Italy}
\affiliation[c]{Istituto Nazionale di Fisica Nucleare, Sezione di Roma 3, Via della Vasca Navale 84, 00146 Rome, Italy}

\author[d,*]{Gianluca Calcagni\,\orcidlink{0000-0003-2631-4588},\note{Corresponding author}}
\emailAdd{g.calcagni@csic.es}
\affiliation[d]{Instituto de Estructura de la Materia, CSIC, Serrano 121, 28006 Madrid, Spain}

\author[e]{Leonardo Modesto\,\orcidlink{0000-0003-2783-8797}}
\emailAdd{leonardo.modesto@unica.it}
\affiliation[e]{Dipartimento di Fisica, Universit\`a di Cagliari, Cittadella Universitaria, 09042 Monserrato, Italy}

\author[f,g]{and Giuseppe Nardelli\,\orcidlink{0000-0002-7416-6332}}
\emailAdd{giuseppe.nardelli@unicatt.it}
\affiliation[f]{Dipartimento di Matematica e Fisica, Universit\`a Cattolica del Sacro Cuore, Via della Garzetta 48, 25133 Brescia, Italy}
\affiliation[g]{TIFPA -- INFN, c/o Dipartimento di Fisica,Universit\`a di Trento, 38123 Povo (TN), Italy}

\abstract{We discuss the conical region of convergence of exponential and asymptotically polynomial form factors and their integral representations. Then, we calculate the spectral representation of the propagator of nonlocal theories with entire form factors, in particular, of the above type. The spectral density is positive-definite and exhibits the same spectrum as the local theory. We also find that the piece of the propagator corresponding to the time-ordered two-point correlation function admits a generalization of the K\"all\'en--Lehmann representation with a standard momentum dependence and a spectral density differing from the local one only in the presence of interactions. These results are in agreement with what already known about the free theory after a field redefinition and about perturbative unitarity of the interacting theory. The spectral and K\"all\'en--Lehmann representations have the same standard local limit, which is recovered smoothly when sending the fundamental length scale $\ell_*$ in the form factor to zero.}


\maketitle
\renewcommand{\thefootnote}{\arabic{footnote}}
%\tableofcontents

%%%%%%%%%%%%%%%%%%%%%%%%%%%%%%%%%%%%%%%%%%%%%%%%%%%%%%%%%%%%%%%%%%%%%%%%%%%%%
%%%%%%%%%%%%%%%%%%%%%%%%%%%%%%%%%%%%%%%%%%%%%%%%%%%%%%%%%%%%%%%%%%%%%%%%%%%%%

\section{Introduction and main result}\label{sec1}

Among the candidates for a theory of quantum gravity, one that has actively been studied in the last 15 years is nonlocal quantum gravity (NLQG), in particular, asymptotically local quantum gravity (ALQG) initiated in \cite{Krasnikov:1987yj,Kuzmin:1989sp,Tomboulis:1997gg,Modesto:2011kw,Biswas:2011ar} (see \cite{BasiBeneito:2022wux,Buoninfante:2022ild,Koshelev:2023elc} for reviews). NLQG is a Lorentz-invariant, diffeomorphism invariant, perturbative quantum field theory of graviton and matter fields whose dynamics is characterized by operators, called form factors, with infinitely many derivatives. The gravitational action is quadratic in curvature operators just like Stelle gravity \cite{Stelle:1976gc,Stelle:1977ry,Julve:1978xn,Fradkin:1981iu,Avramidi:1985ki,Hindawi:1995uk} and, for this reason, it is renormalizable. The presence of nonlocal form factors of a certain type, inserted between each pair of curvature tensors $R_{\mu\nu\s\tau}$, $R_{\mu\nu}$ and $R$, further improves the ultraviolet (UV) behaviour of the theory and also makes it ghost free, thus overcoming the unitarity problem of Stelle gravity.

Despite much progress in the formulation of the theory, there are still some elementary points of confusion that deserve an in-depth treatment. One such point is about the structure of the propagator in relation with unitarity. Ignoring tensorial indices, consider the typical massive scalar propagator (or, more precisely, the Green's function, which is $-\rmi$ times the propagator) appearing in ALQG. In momentum space and at the tree level, it is
\be\label{prop}
\tilde G(-k^2)=\frac{\rme^{-\H(-k^2-m^2)}}{k^2+m^2}\,,
\ee
where we work in $(-,+,+,+)$ signature, $-k^2=k_0^2-|\bm{k}|^2$ is the square of the momentum, the Feynman prescription $k^2\to k^2-\rmi\e$ is understood in the denominator and $\exp\H(z)$ is an entire function such that
\be\label{Hir}
\H(0)=0
\ee
on-shell or in the infrared (IR) in the massless case and that does not introduce any extra pole with respect to the standard case $\tilde G(-k^2)=1/k^2$. For the graviton and other massless fields, $m=0$. The form factors employed in ALQG are asymptotically polynomial, meaning that in the UV limit they scale as $\exp\H(z)\sim z^n$, i.e.,
\be\label{hdprop}
\tilde G(-k^2)\sim \frac{1}{k^{2n+2}}\,,\qquad n\in\mathbbm{N}^+\,.
\ee
If the theory is unitary, as shown in \cite{Krasnikov:1987yj,Kuzmin:1989sp,Tomboulis:1997gg,Modesto:2011kw,Biswas:2011ar,Briscese:2018oyx}, then standard arguments of quantum field theory would suggest that the propagator admits a standard Källén--Lehmann representation and that the theory admits a Hamiltonian description. Let us recall that, for a Lorentz-invariant Green's function $\tilde G(-k^2)$, the standard Källén--Lehmann representation is \cite{Kal52,Leh54,tV74b,Sre07,Sch14,Zwi16}
\be\label{kale}
\tilde G(-k^2)=\int_{0}^{+\infty}\rmd s\,\frac{\rho(s)}{s+k^2-\rmi\e}\,,
\ee
where $s$ is a real positive parameter and $\rho(s)$ is called spectral function or spectral density. In particular, Weinberg \cite[section 10.7]{Wei95} argued that the left-hand side of \Eq{kale} cannot scale faster than $k^{-2}$ in the UV if $\rho(s)\geq 0$ for all $s$, the condition for unitarity. Two different proofs of that are given in \cite{Sch14} and \cite[appendix E.3]{Calcagni:2022shb}. This theorem has a strong implication for alternative quantum field theories, including of gravity: higher-derivative theories cannot be unitary because their propagator typically scales as \Eq{hdprop} with $n>2$. But then, precisely for the same reason, should not also ALQG be ruled out? Then what to make of unitarity claims? The casual reader might find this query into NLQG non-trivial.

In this paper, we will answer to the above question in the case of a scalar field; the case of gravity only differs in the tensorial structure and does not add anything to the main point made here, which is the following. As is already known, micro-causality is violated in interacting nonlocal theories and the time ordering in scattering amplitudes produces an extremely complicated result, which we will recall in section \ref{Tprod}. This non-causality at microscopic scales of order of the Planck length around the light cone does not forbid the theory to have a well-defined complete basis of on-shell Fock states. It is on these states that time ordering of nonlocal fields makes sense. This situation results in a mismatch between the \emph{diagrammatic} propagator $\tilde G(-k^2)$, or simply \emph{the} propagator, appearing in the calculation of Feynman amplitudes, and its \emph{time-ordered} part $\tilde G_{\rm to}(-k^2)$. In the former case, the contour prescription in $k^0$ cannot be interpreted as a time ordering due to the presence of the form factor that forbids to close such contour at infinity as in the local case. In other words, while in the standard local case the left-hand side of \Eq{kale} is the Fourier transform of the time-ordered two-point correlation function, in nonlocal theories it is not. Hence, the Källén--Lehmann representation, which is the integral representation of the time-ordered propagator, is inequivalent to the integral representation of the diagrammatic propagator.
	 
It turns out that the diagrammatic propagator of ALQG admits a generalized integral representation different from \Eq{kale} but with a similar structure where the momentum-dependent part is dressed, at tree level in perturbation theory, by a form factor:
\be\label{intfin}
\boxd{\tilde G(-k^2)=\int_{0}^{+\infty}\rmd s\,\frac{\rme^{-\H(-k^2-s)}}{s+k^2-\rmi\e}\,\rho_{\rm tree}(s)\,,\qquad \rho_{\rm tree}(s)=\de(s-m^2)\,,}
\ee
which obviously gives the correct result \Eq{prop}. The extra form-factor term is not included in the definition of the spectral density because it is entire, hence it does not affect the physical spectrum. In the presence of interactions,
\be\label{dialocint}
\boxd{\!\!\!\tilde G(-k^2)=\int_{0}^{+\infty}\rmd s\,\frac{\rme^{-\H(-k^2-m^2)}}{s+k^2-\rmi\e}\,\rho(s)\,,\qquad\! \rho(s)=
\sum_\la \de(s-m_\lambda^2) \, \vert\langle\Omega\vert\tilde \phi(0)  \vert \la_0 \rangle\vert^2,\!\!\!}
\ee
where $\tilde\phi(0)=\rme^{\H(\B)/2}\phi(x)|_{x=0}$, $|\Om\rangle$ is the vacuum state and the sum is on a complete set of on-shell states $|\la_0\rangle$ of zero spatial momentum and mass $m_\la$. Note that \Eq{dialocint} implies \Eq{intfin}. Other representations are possible but they obscure information on the physical spectrum of the theory. We also calculate the Källén--Lehmann representation for the contribution in the propagator corresponding to the time-ordered two-point function, which in momentum space for the free case (i.e., without interactions) is
\be\label{kalefin}
\boxd{\tilde G_{\rm to}(-k^2)=\int_{0}^{+\infty}\rmd s\,\frac{\rho_{\rm free}(s)}{s+k^2-\rmi\e}\,,\qquad \rho_{\rm free}(s)=\de(s-m^2)\,,}%=\rme^{-\H(s-m^2)}\de(s-m^2)
\ee
where ``to'' stands for time-ordered, while in the presence of interactions it is
\be\label{kalefinint}
\boxd{\tilde G_{\rm to}(-k^2)=\int_{0}^{+\infty}\rmd s\,\frac{\rho(s)}{s+k^2-\rmi\e}\,,\qquad \rho(s)	=  \rme^{-\H(s-m^2)}\sum_\la \delta(s-m_\lambda^2) \, \vert\langle\Omega\vert\tilde \phi(0)  \vert \la_0 \rangle\vert^2,}%=\rme^{-\H(s-m^2)}\de(s-m^2)
\ee
which implies \Eq{kalefin}. The spectral function is positive-definite, thus confirming what we already know from a field redefinition (section \ref{sec2a}): nonlocal theories with entire form factors are free-level unitary. We will come to understand how ALQG can be unitary and evade Weinberg's theorem without contradiction. The key, in action also in another NLQG called fractional quantum gravity \cite{Calcagni:2022shb}, is that the standard representation \Eq{kale} is not valid, thus violating one of the hypotheses of the theorem.

All these results are tightly related to the mathematical and physical properties of entire form factors, on which we will gain insight. We will find that the Cauchy representation of the form factors and of the propagator are valid only in a certain conical region $\cC$ where the contribution of arcs at infinity vanishes. This is perhaps the clearest way to understand the origin of such a region, not only for asymptotically polynomial form factors \cite{Kuzmin:1989sp,Tomboulis:1997gg} but also for exponential form factors.
 
The rest of the paper contains the derivation of \Eq{intfin}--\Eq{kalefinint} and discusses in detail, for the first time, the domain of convergence of the form factors as well as the physical interpretation of the above and of a different integral representation of the propagator as the superposition of complex conjugate modes. 

In section \ref{sec2}, we derive the integral representation of the Green's function when the latter is expressed as a Cauchy integral on a specific type of contour. The example of the standard local case is discussed in section \ref{sec2a}. In section \ref{sec3}, we turn to nonlocal form factors. After showing that in NLQG \Eqq{kale} cannot hold and that Weinberg's theorem on unitarity is violated, in section \ref{sec3a}, we recall that a well-known nonlocal field redefinition would indicate the contrary at the tree level. This conundrum is solved in section \ref{sec3c} but, before that, in section \ref{sec3b} we study in detail the exponential and asymptotically polynomial form factors appearing in fundamentally nonlocal field theories, with particular emphasis on their Cauchy representations and domain of convergence. In particular, we give a comprehensive, step-by-step description of the conical region where these form factors are defined. The derivation of the generalized representations \Eq{intfin} and \Eq{dialocint} in nonlocal theories with such form factors and their physical interpretation are presented in section \ref{sec3c}. In section \ref{Tprod}, we study the relation between time ordering and the two-point correlation function in NLQG and find the generalized Källén--Lehmann representation with and without interactions (eqs.\ \Eq{kalefinint} and \Eq{kalefin}, respectively). Conclusions are in section \ref{sec4}. Appendix \ref{appA} is not necessary at a first reading but it illustrates that a choice of contour not corresponding to the spectral representation leads to an expression of the form factor or of the propagator where information on the physical spectrum is not apparent.

%%%%%%%%%%%%%%%%%%%%%%%%%%%%%%%%%%%%%%%%%%%%%%%%%%%%%%%%%%%%%%%%%%%%%%%%%%%%%
%%%%%%%%%%%%%%%%%%%%%%%%%%%%%%%%%%%%%%%%%%%%%%%%%%%%%%%%%%%%%%%%%%%%%%%%%%%%%

\section{Cauchy integral}\label{sec2}

The standard representation \Eq{kale} is the integral over the real half-line of a propagator-like term $1/(s+k^2)$ weighted by a spectral function $\rho(s)$. The integration parameter $s$ plays the role of a squared mass, so that one is summing over mass modes weighted by the spectral density of the physical states. An extremely convenient way to derive it, which does not assume anything about the underlying theory except for Lorentz invariance, is via the Cauchy representation of the Green's function.

Consider the Fourier transform $\tilde G(-k^2)$ of the Green's function $G$ in a Lorentz-invariant theory and regard it as a function $\tilde G(z)$ on the complex plane. Assume that $\tilde G(z^*)=\tilde G^*(z)$, i.e., that it is real on the real axis. Given a counter-clockwise contour $\G$ encircling the point $z=-k^2$ and such that $\tilde G(z)$ is holomorphic (i.e., analytic) inside and on $\G$, Cauchy's integral representation of the Green's function is
\be\label{opt}
\tilde G(-k^2)=\frac{1}{2\pi\rmi}\oint_{\G}\rmd z\,\frac{\tilde G(z)}{z+k^2}\,,
\ee
where from now on we ignore the Feyman prescription $k^2\to k^2-\rmi\e$. If $\tilde G(z)$ has poles and branch cuts on the complex plane, then the contour $\G$ is chosen so that to keep all such singularities outside; deformations of this contour not crossing the singularities are homotopic, hence all mathematically and physically equivalent.

At infinity in the complex plane, the contour $\G$ is made of one or more arcs of radius $R$, whose contribution must be zero in the limit $R\to+\infty$. Around poles and branch points, one can deform $\G$ as a small circle of radius $\ve$ and then send $\ve\to 0^+$; in this limit, the contribution of poles and branch points must be finite. Along branch cuts, $\G$ can be deformed as a pair of opposite semi-lines originating or ending at the branch point and running along the cut on either side \cite{Calcagni:2022shb}. The net result of this computation, that is, the contributions of each pole and branch cut can be written as an integral over the real variable $s\geq 0$ on a semi-finite interval:
\be\label{kalegen}
\tilde G(-k^2)=\int_0^{+\infty}\rmd s\,\cI(s,k)\,.
\ee
We call this expression \emph{generalized spectral representation} if the contour $\G$ is kept squeezed along the real line and straight when passing near the poles, either because it runs through the real line and poles thereon are displaced infinitesimally or because the path runs parallel to the real line by an infinitesimal imaginary displacement. If, instead, the contour is deformed to pass around such points, \Eq{kalegen} is not, in general, a spectral representation, as we will see below. Finally, if we replace the left-hand side of \Eq{kalegen} with the time-ordered Green's function $\tilde G_{\rm to}(-k^2)$ we call this \emph{generalized Källén--Lehmann representation}. In the most general nonlocal case, ``spectral'' and ``Källén--Lehmann'' are not synonyms, since $\tilde G(-k^2)\neq \tilde G_{\rm to}(-k^2)$.\footnote{In \cite{Calcagni:2022shb,Calcagni:2021ljs}, the spectral representation was improperly called Källén--Lehmann.}

%%%%%%%%%%%%%%%%%%%%%%%%%%%%%%%%%%%%%%%%%%%%%%%%%%%%%%%%%%%%%%%%%%%%%%%%%%%%%

\subsection{Standard spectral representation}\label{sec2a}

The most important thing to remind about the contour $\G$ defining the spectral (and, in the local case, also the Källén--Lehmann) representation is to keep it straight near poles. The local case 
\be\label{Gstd}
\tilde G(-k^2)=\frac{1}{k^2+m^2}\,,
\ee
can teach us a lot about the choice of contour $\G$ as well as about the spectral representation in NLQG. Physically and mathematically, when we have a pole on the real line it is equivalent to deform the contour around the pole, to shift the pole to imaginary coordinates or to shift the contour instead:
\ba
\parbox{6cm}{\includegraphics[width=6cm]{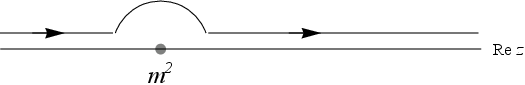}} &=&\quad \parbox{6cm}{\includegraphics[width=6cm]{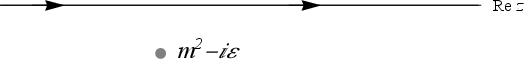}}\nn\nn
&=& \quad\parbox{6cm}{\includegraphics[width=6cm]{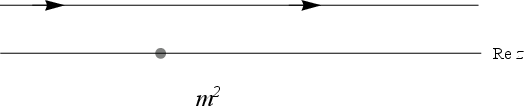}}.\label{ppp}
\ea
However, in the first case (left-hand side) one integrates over an arc around the pole of radius $\ve$ which corresponds to a fraction of the residue; it is identically equal to the left-hand side of \Eq{opt}. In the second case (right-hand side), one is forcing the integration on the real line. It is the latter that corresponds to the spectral representation. In particular, the Cauchy representation \Eq{opt} corresponds to the second line in \Eq{ppp}.

In the local case \Eq{Gstd}, the contour $\G'$ corresponding to \Eq{Gstd} (i.e., the left-hand side of \Eq{kale}) and the contour $\G$ corresponding to the right-hand side of \Eq{kale} are
\be
\parbox{6.5cm}{\includegraphics[width=6.5cm]{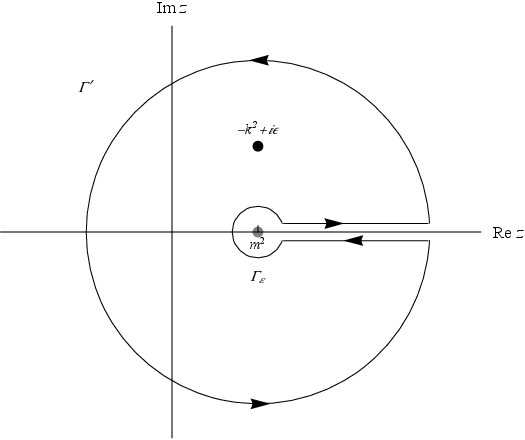}} = \parbox{6.5cm}{\includegraphics[width=6.5cm]{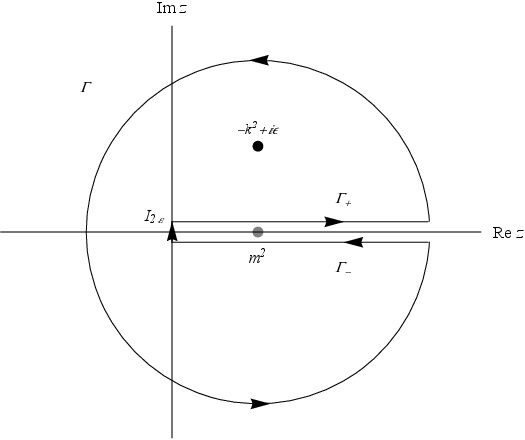}}.\label{equi}
\ee
To see this, if we compute the Cauchy integral on $\G'$, we can throw away both the arc at infinity (whose contribution scales as $1/R$ and the radius $R$ is sent to infinity; more on this below) and the integration along the positive semi-axis, where the function
\be\label{Gzstd}
\tilde G(z)=\frac{1}{m^2-z}
\ee
is holomorphic and there is no discontinuity. The only surviving contribution is around the arc $\G_\ve$ parametrized by $z=m^2+\ve\,\exp(\rmi\t)$:
\ba
\frac{1}{2\pi\rmi}\oint_{\G'}\rmd z\,\frac{\tilde G(z)}{z+k^2}&=& \frac{1}{2\pi\rmi}\int_{\G_\ve}\rmd z\,\frac{\tilde G(z)}{z+k^2}\nn
&=&\lim_{\ve\to 0^+} \frac{\rmi\ve}{2\pi\rmi}\int_{2\pi}^0\rmd \t\,\rme^{\rmi\t}\frac{1}{m^2+\ve\rme^{\rmi\t}+k^2}\frac{1}{(-\ve)\rme^{\rmi\t}}\nn
&=& \lim_{\ve\to 0^+}-\frac{1}{2\pi}\frac{1}{k^2+m^2}\int_{2\pi}^0\rmd \t+O(\ve)\nn
&=& \frac{1}{k^2+m^2}\,.\label{lhs}
\ea
On the other hand, the contour $\G=I_{2\ve}\cup\G_+\cup\G_-$ is made of an arc at infinity (which gives a zero contribution as before), a small vertical segment $I_{2\ve}$ of length $2\ve$ parametrized by $z=\rmi t$ and corresponding to a vanishing integral,
\be\label{Ive}
\frac{1}{2\pi\rmi}\int_{I_{2\ve}}\rmd z\,\frac{\tilde G(z)}{z+k^2}= \lim_{\ve\to 0^+}\frac{\rmi}{2\pi\rmi}\int_{-\ve}^{\ve}\rmd t\,\frac{1}{\rmi t+k^2}\frac{1}{m^2-\rmi t}\simeq\lim_{\ve\to 0^+}\frac{\ve}{\pi k^2m^2}=0\,,
\ee
plus the contribution of the integration on $\G_+\cup\G_-$ parallel to the real line,
\ba
\frac{1}{2\pi\rmi}\int_{\G}\rmd z\,\frac{\tilde G(z)}{z+k^2}&=&
\frac{1}{2\pi\rmi}\int_{\G_+\cup\G_-}\rmd z\,\frac{\tilde G(z)}{z+k^2}\nn
&=&\lim_{\ve\to 0^+}\int_0^{+\infty}\rmd s\left[\frac{\tilde G(s+\rmi\ve)}{s+\rmi\ve+k^2}-\frac{\tilde G(s-\rmi\ve)}{s-\rmi\ve+k^2}\right]\nn
&=&\lim_{\ve\to 0^+}\frac{1}{2\pi\rmi}\int_0^{+\infty}\rmd s\left[\frac{1}{m^2-s-\rmi\ve}\frac{1}{s+\rmi\ve+k^2}\right.\nn
&&\qquad\qquad\qquad\qquad\left.-\frac{1}{m^2-s+\rmi\ve}\frac{1}{s-\rmi\ve+k^2}\right]\nn
&=&\lim_{\ve\to 0^+}\frac{1}{2\pi\rmi}\int_0^{+\infty}\rmd s\,\frac{1}{s+k^2}\left[\frac{1}{m^2-s-\rmi\ve}-\frac{1}{m^2-s+\rmi\ve}\right]+O(\ve)\nn
&=&\lim_{\ve\to 0^+}\frac{1}{2\pi\rmi}\int_0^{+\infty}\rmd s\,\frac{1}{s+k^2}\left[\frac{1}{m^2-s-\rmi\ve}\frac{m^2-s+\rmi\ve}{m^2-s+\rmi\ve}\right.\nn
&&\qquad\qquad\qquad\qquad\qquad\quad\left.-\frac{1}{m^2-s+\rmi\ve}\frac{m^2-s-\rmi\ve}{m^2-s-\rmi\ve}\right]\nn
&=&\lim_{\ve\to 0^+}\frac{1}{2\pi\rmi}\int_0^{+\infty}\rmd s\,\frac{1}{s+k^2}\left[\frac{m^2-s}{(m^2-s)^2+\ve^2}+\frac{\rmi\ve}{(m^2-s)^2+\ve^2}\right.\nn
&&\qquad\qquad\qquad\qquad\qquad\quad\left.-\frac{m^2-s}{(m^2-s)^2+\ve^2}+\frac{\rmi\ve}{(m^2-s)^2+\ve^2}\right]\nn
&=& \int_0^{+\infty}\rmd s\,\frac{\de(s-m^2)}{s+k^2}\,,\label{rhs}%\stackrel{\textrm{\tiny \Eq{plemsok}}}{=}
\ea
where in the last step we used the Sokhotski--Plemelj formula
\be
\frac{1}{x-\rmi\e}=\frac{1}{x-\rmi\e}\,\frac{x+\rmi\e}{x+\rmi\e}=\frac{x}{x^2+\e^2}+\rmi\,\frac{\e}{x^2+\e^2}\stackrel{\e\to 0^+}{=} {\rm PV}\!\left(\frac{1}{x}\right)+\rmi\pi\de(x)\,,\label{plemsok}
\ee
and PV is the principal value.

Since the two contours $\G'$ and $\G$ are homotopic, \Eq{lhs} and \Eq{rhs} are equal to each other and the standard Källén--Lehmann representation \Eq{kale} with spectral density
\be\label{rstd}
\rho(s)=\de(s-m^2)
\ee
is proven. The representation \Eq{kale} with a different $\rho(s)$ actually holds not only in the presence of interactions, but also for \emph{any} interactive theory admitting a time-ordered product \cite[section 24.2.1]{Sch14}. Not even locality is assumed. Thanks to its generality, this result is very powerful and, as said above, it severely constrains the propagator of unitary theories. Thus, if the standard spectral (or Källén--Lehmann, since $\tilde G=\tilde G_{\rm to}$ in this case) representation \Eq{kale} holds, then Weinberg's theorem holds.

A final observation on the local propagator will turn out to be very useful in the nonlocal case. The propagator can be represented not only with the Cauchy representation \Eq{opt} evaluated on different contours, but also with an altogether different Cauchy representation where the contour $\tilde\G$ now is clockwise and circumscribes only the $z=m^2$ pole:
\be\label{optm}
\tilde G(-k^2)=\frac{1}{2\pi\rmi}\oint_{\tilde\G}\rmd z\,\frac{\tilde G(z,k^2)}{z-m^2}\,,
\ee
where in the local case
\be
\tilde G(z,k^2)=\frac{1}{z+k^2}\,.
\ee
Clearly, this is a naive rewriting of \Eq{opt} with propagator \Eq{Gzstd} but with a contour $\tilde\G$ where the poles $z=m^2$ and $z=-k^2$ exchange roles and the latter (now pole of $\tilde G$) is left outside the contour. Then, the equivalence between \Eq{opt} and \Eq{optm} can be graphically expressed as
\be
\parbox{6.5cm}{\includegraphics[width=6.5cm]{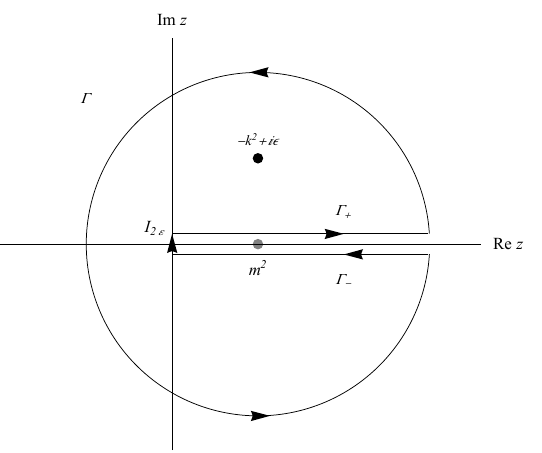}} = \parbox{6.5cm}{\includegraphics[width=6.5cm]{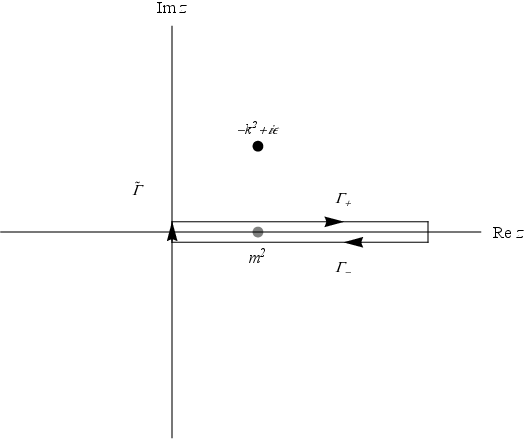}}.\label{equi2}
\ee
The contribution $I_{2\ve}$ of the left short side of the contour is zero according to \Eq{Ive} and so is the one on the right side $I_{\ve,R}$:
\ba
\frac{1}{2\pi\rmi}\int_{I_{\ve,R}}\rmd z\,\frac{\tilde G(z)}{z+k^2}&\stackrel{\textrm{\tiny $z=R+\rmi t$}}{=}& \int_{-\ve}^{\ve}\rmd t\,\frac{1}{R+\rmi t+k^2}\frac{1}{m^2-R-\rmi t}\nn
&\simeq& \frac{\ve}{\pi(R+k^2)(m^2-R)}\to 0\,,\label{IveR}
\ea
where the ordering of the limits $\ve\to 0^+$ and $R\to +\infty$ does not matter. Therefore, the only non-zero contribution to the contour is identical step by step to the one in \Eq{rhs}.

%%%%%%%%%%%%%%%%%%%%%%%%%%%%%%%%%%%%%%%%%%%%%%%%%%%%%%%%%%%%%%%%%%%%%%%%%%%%%
%%%%%%%%%%%%%%%%%%%%%%%%%%%%%%%%%%%%%%%%%%%%%%%%%%%%%%%%%%%%%%%%%%%%%%%%%%%%%

\section{Nonlocal form factors}\label{sec3}

The propagator \Eq{prop} is a tree-level expression that, according to the results on renormalizability in ALQG \cite{Modesto:2014lga,Modesto:2015lna,Calcagni:2023goc} (reviewed in \cite{BasiBeneito:2022wux}), receives quantum corrections that modify the structure of the form factor $\exp(-\H)$. However, loop corrections to the propagator do not imply a spoiling of the unitarity property, as has been checked at all perturbative orders \cite{Briscese:2018oyx,Efimov:1967dpd,AE1,Pius:2016jsl}; instrumental to this result is the fact that the function appearing in the numerator of \Eq{prop} is entire. This form factor is the fundamental building block with which to construct Feynman diagrams at higher loop orders and it is thus indispensable to study its basic properties.

In nonlocal theories with propagator \Eq{prop}, Weinberg's theorem does not apply because the Källén--Lehmann representation \Eq{kale} does not hold and the latter does not hold because, for the full propagator, the time-ordered product is deformed by the form factors, a situation not contemplated in the derivation of \cite{Sch14}. A very simple way to see this is based upon the fact that the perturbative spectrum of the theory is the same as in a local theory. If there were a standard spectral representation, then there would exist a spectral density $\rho(s)$ such that (take $m=0$ for simplicity)
\be\label{counter1}
\frac{\rme^{-\H(-k^2)}}{k^2-\rmi\e}\stackrel{\textrm{?}}{=}\int_0^{+\infty}\rmd s\,\frac{\rho(s)}{s+k^2-\rmi\e}\,.
\ee
Using the Sokhotski--Plemelj formula \Eq{plemsok} and taking the imaginary part of both sides of \Eqq{counter1}, one gets
\be
\hspace{-.2cm}\rmi\pi\de(k^2)\,\rme^{-\H(-k^2)} =\rmi\pi \int_0^{+\infty}\!\rmd s\,\rho(s)\,\de(s+k^2)=\rmi\pi\rho(-k^2)\qquad \Longrightarrow\qquad \rho(k^2)=\de(k^2),
\ee
where we used \Eq{Hir}. Thus, the spectral density is the same as in the local theory and the spectrum only has a spin-0 point particle with dispersion relation $k^2=0$. This conclusion is physically correct but it leads to a contradiction in \Eqq{counter1}, since the form factor $\exp\H$ is non-trivial. Therefore, \Eq{counter1} cannot be valid.

%%%%%%%%%%%%%%%%%%%%%%%%%%%%%%%%%%%%%%%%%%%%%%%%%%%%%%%%%%%%%%%%%%%%%%%%%%%%%
	
\subsection{Definitions of nonlocal operators}\label{sec30}

In this sub-section, for the reader's convenience we review two common definitions of operators with infinitely many derivatives. The expert can skip this material since it is not needed for the rest of the paper.

The natural class of functions on which to define the nonlocal operators appearing in NLQG is the space $\cS$ of rapidly decreasing $C^\infty$ functions along all spacetime directions. In the topology of the space $L^2$ of square-integrable functions, $\cS$ is dense in $L^2$. As an example, the action of an exponential operator on a Gaussian function is calculated in \cite[section 2.4.1]{BasiBeneito:2022wux}. Clearly, the operator $\rme^{\H(\B)}$ is invertible in such a functional space, as can be easily seen in Fourier transform. In general, however, nonlocal form factors can be defined on a larger functional space depending on the type of representation we employ for them. As recalled in \cite[section 2.2]{BasiBeneito:2022wux}, there are two ways to represent nonlocal operators. One is as a series
\be\label{serie}
\rme^{\H(\B)}=\sum_{n=1}^\infty a_n\, \B^n
\ee
with certain coefficients $a_n$.\footnote{In general, but not in the case of NLQG, this expression can be only formal (for instance, operators such as $\sqrt{\B}$ are first regularized and then represented as a series with some divergent coefficients $a_n$) or extended to a Laurent series.} The other is via an integral kernel $K$
\be\label{inter}
\rme^{\H(\B)}\phi(x)=\int \rmd^D y\,K(y-x)\,\phi(y)\,,
\ee
which is valid on Minkowski spacetime but can be generalized to any curved background \cite{Calcagni:2007ru} (this procedure is reviewed in \cite[section 2.3.2]{BasiBeneito:2022wux}). It turns out that the integral representation is more general than the series one because it can be well-defined even on functions which are not in $L^2$. An example is given in \cite{Calcagni:2007ru} for a cosmological background with power-law scale factor. The function $f(t)=t^p$ is not square-integrable on the real line and the operator $\rme^\B=\sum_n\B^n/n!$ represented as a series \Eq{serie} does not converge if $p$ is non-integer, since the terms $\B^n t^p$ become progressively large as $n$ increases, where $\B=-\p_t^2-3H\p_t$ and $H\propto 1/t$. Remarkably, however, in the integral representation the object $\rme^\B t^p$ is a well-defined function not in $L^2$. This observation culminated in the formulation of the diffusion method as a way to find solutions of nonlocal systems \cite{Calcagni:2007ru,Mulryne:2008iq,Calcagni:2009tx,Calcagni:2009jb,Calcagni:2018lyd}, including NLQG \cite{Calcagni:2018gke,Calcagni:2018fid}. It implied a radical advance with respect to early attempts to find solutions of nonlocal systems using the series representation, which failed in their majority because they led to an apparently paradoxical Cauchy problem of initial conditions \cite{Moeller:2002vx}. The main issue with the series representation is that, if truncated, it leads to a higher-derivative model which is not an analytic approximation of the original nonlocal one and has a different physical spectrum.

The equations of motion for the matter and gravitational sector of nonlocal field theories have been computed several times, both with the series and the integral definitions \cite{Biswas:2011ar,Calcagni:2018lyd,Calcagni:2018gke,Koshelev:2013lfm,Conroy:2014eja,Briscese:2015urz,Teimouri:2016ulk}.

%%%%%%%%%%%%%%%%%%%%%%%%%%%%%%%%%%%%%%%%%%%%%%%%%%%%%%%%%%%%%%%%%%%%%%%%%%%%%

\subsection{Field redefinition in nonlocal theories}\label{sec3a}

A very well-known trick can quickly show that nonlocal theories with an entire form factor can be recast in a form that respects the representation \Eq{kale} and, therefore, also Weinberg's theorem. Consider the nonlocal massless scalar Lagrangian
\be\label{lag1}
\cL=\frac12\phi\B\,\rme^{\H(\B)}\phi-V(\phi)\,,
\ee
where $\B$ is the Laplace--Beltrami operator ($\B\to-k^2$ after Fourier transforming on Min\-kow\-ski spacetime) and $V$ is a local potential. %Note that in momentum space $\H(\B)\to\H(-k^2)$ and that we systematically omit the $-$ sign in the argument of $\H(k^2)$.
 If $\exp\H$ is entire, then the field redefinition
\be\label{fire}
\tilde\phi\coloneqq \rme^{\frac12\H(\B)}\phi
\ee
does not change the spectrum of the theory, so that one can recast \Eq{lag1} as
\be\label{lag2}
\cL=\frac12\tilde\phi\B\,\tilde\phi-V\left[\rme^{-\frac12\H(\B)}\tilde\phi\right]+\dots\,,
\ee
up to total derivatives. Therefore, for a free field ($V=0$), the nonlocal dynamics $0=\de\cL/\de\phi=\B\,\rme^\H\phi$ is equivalent to the local dynamics $0=\de\cL/\de\tilde\phi=\B\tilde\phi$, $\tilde\phi$ has a standard kinetic term and the spectral representation of its Green's function is the standard one, \Eqq{kale}.

If this were the end of the story, there would have been no need of this article. However, as soon as one switches on interactions, the field redefinition \Eq{fire} only displaces nonlocality from the kinetic term to the potential, so that the $\tilde\phi$ system is still nonlocal. Then, we expect nonlocality to pop in again at the first and higher orders in the loop expansion and to modify the spectral representation \Eq{kale} accordingly. This, again, should be consistent with known results on perturbative unitarity \cite{Briscese:2018oyx}. Thus, we conclude that the field redefinition clarifies the issue posed in the introduction only at the free level (tree level and in the absence of interactions), while leaving it unanswered at one- and higher-loop orders.

%%%%%%%%%%%%%%%%%%%%%%%%%%%%%%%%%%%%%%%%%%%%%%%%%%%%%%%%%%%%%%%%%%%%%%%%%%%%%

\subsection{Cauchy representations of nonlocal form factors}\label{sec3b}

Just like the propagator, nonlocal form factors admit a class of Cauchy representations labelled by the choice of contour $\G$:
\be
\rme^{-\H(-k^2-m^2)}=\frac{1}{2\pi\rmi}\oint_{\G}\rmd z\,\frac{\rme^{-\H(z-m^2)}}{z+k^2}\,,\qquad z\in\cC\,.\label{fofamain}
\ee
We discuss these representations here for the first time because, on one hand, they provide a transparent derivation of the conical region in the complex plane on which such operators are well-defined and, on the other hand, they elucidate how different choices of contour $\G$ can shed light on different physical aspects associated to these form factors. In \Eq{fofamain}, we have chosen to represent $\exp(-\H)$ instead of $\exp(+\H)$ because the former is the quantity appearing in Feynman diagrams; the convergence properties of $\exp(+\H)$ can be immediately inferred from the former and will be discussed later.

We consider two types of form factors $\exp\H$:
\begin{enumerate}
	\item Exponential $\exp\H$ (monomial $\H$):
	\be\label{H1}
	\H(z)=(-z)^n\,,\qquad n=1,2,3,\dots\,.
	\ee
	This type includes the form factors $\exp\H=\exp(-\B)$, $\exp\H=\exp(\B^2)$, which were proposed by Wataghin \cite{Wataghin:1934ann} and Krasnikov \cite{Krasnikov:1987yj}, respectively.
	\item Asymptotically polynomial $\exp\H$ (special function $\H$):
	\be
	\H(z)={\rm Ein}[(-z)^n]\coloneqq\int_0^{(-z)^n}\!\rmd\om\,\frac{1-\rme^{-\om}}{\om}=\ln(-z)^n+\G[0,(-z)^n]+\g_{\rm E}\,,\label{H2}
	\ee
	where Ein is the complementary exponential integral \cite[formula 6.2.3]{NIST}, $\G$ is the upper incomplete gamma function \cite[formula 8.2.2]{NIST} and $\g_{\rm E}$ is Euler--Mascheroni's constant.\footnote{Taking the branch cut of the incomplete gamma on the negative real axis, the last equality in \Eq{H2} holds for $-\pi<\arg (-z)^n<\pi$ (hence $\pi(-n^{-1}-1)<\arg z<\pi(n^{-1}-1)$, where we used $-1=\exp(\rmi\pi)$). However, it can be extended by analytic continuation elsewhere on the complex plane.} This type of form factor is the one considered by Kuz'min and Tomboulis \cite{Kuzmin:1989sp,Tomboulis:1997gg}, $\exp\H=(-\B)^n\exp\{\G[0,(-\B)^n]+\g_{\rm E}\}$, and has the asymptotic limit corresponding to \Eq{hdprop},
	\be
	\rme^{\H(z)}\simeq \rme^{\g_{\rm E}}(-z)^n\,,\qquad z\in\cC\,.
	\ee
\end{enumerate}
For both classes of form factors, we employ dimensionless units where the argument of $\H$ hides a fundamental length scale $\lst$ which is part of the definition of the theory and is expected to be of order of the Planck scale $\lpl$. Therefore, in what follows, the argument of $\H$ should be understood as $\H(\B)= \H(\lst^2\B)$ in position space and $\H(-k^2)= \H(-\lst^2k^2)$ in momentum space. Expressions with a mass $m$ are generalized accordingly as $\H(\B-m^2)= \H[\lst^2(\B-m^2)]$ and $\H(-k^2-m^2)= \H[-\lst^2(k^2+m^2)]$, respectively. Expressions in the complex plane have a dimensionless variable $z$ or $s$ and a rescaled momentum or mass, so that $z+k^2\to z+\lst^2k^2$ and $z-m^2\to z-\lst^2m^2$ in all formul\ae\ below.

%%%%%%%%%%%%%%%%%%%%%%%%%%%%%%%%%%%%%%%%%%%%%%%%%%%%%%%%%%%%%%%%%%%%%%%%%%%%%

\subsubsection{Arcs at infinity}\label{sec:arcsinf}

Any contour $\G$ in the $(\Re\,z,\Im\,z)$ complex plane chosen in \Eq{fofamain} will be closed at infinity by one or more arcs $\G_R^m$, where $R\to+\infty$ is the radius of the arcs and $m=0,1,\dots,n-1$. While in the standard case there is only one such arc ($n=1$) with opening angle $2\pi$, for exponential and asymptotically polynomial form factors there may be many, all with a non-trivial opening.

Let $\G_R$ be any of the arcs $\G_R^m$ of radius $R$ and opening angle $\t^-<\t<\t^+$ centered at the origin in the $(\Re\,z,\Im\,z)$ plane, parametrized by $z=R\,\exp(\rmi\t)$. We want to find the most general limiting angles $\t^-$ and $\t^+$ such that the contribution of $\G_R$ to \Eq{fofamain} vanishes in the limit $R\to+\infty$. Therefore, we calculate
\ba
I_R&\coloneqq&-\frac{1}{2\pi\rmi}\int_{\G_R}\rmd z\,\frac{\rme^{-\H(z-m^2)}}{z+k^2}= -\frac{R}{2\pi}\int_{\t^-}^{\t^+}\rmd\t\,\rme^{\rmi\t}\frac{\rme^{-\H(R\,\rme^{\rmi\t}-m^2)}}{R\,\rme^{\rmi\t}+k^2}\nn
&=& -\frac{1}{2\pi}\int_{\t^-}^{\t^+}\rmd\t\,\rme^{-\H(R\,\rme^{\rmi\t})}+O(R^{-1})\,,\label{IRgen}
\ea
where we expanded for large $R$ and kept the mass finite. To proceed further, we have to choose form factor. 

For the exponential form factors \Eq{H1} (monomial $\H$), one gets $\H(z)=(-R\,\rme^{\rmi\t})^n=R^n\rme^{\rmi n (\t+\pi)}$ and
\ba
I_R &=& -\frac{1}{2\pi}\int_{\t^-}^{\t^+}\rmd\t\,\rme^{-R^n\rme^{\rmi n (\t+\pi)}}+O(R^{-1})\nn
&=& -\frac{1}{2\pi}\frac{1}{\rmi n}\int_{t(\t^-)}^{t(\t^+)}\rmd t\,\frac{\rme^{-t}}{t}+O(R^{-1})\nn
&=& -\frac{1}{2\pi}\frac{1}{\rmi n}\G(0,t)\big|_{t=t(\t^-)}^{t=t(\t^+)}+O(R^{-1})\nn
&=& -\frac{1}{2\pi}\frac{1}{\rmi n}\G\!\left[0,R^n\rme^{\rmi n (\t+\pi)}\right]\!\bigg|_{\t=\t^-}^{\t=\t^+}+O(R^{-1})\nn
&\simeq& -\frac{1}{2\pi}\frac{1}{\rmi n}\frac{\rme^{-R^n\rme^{\rmi n (\t+\pi)}}}{R^n\rme^{\rmi n (\t+\pi)}}\bigg|_{\t=\t^-}^{\t=\t^+}\,,\label{Eiint}
\ea
where we used $-1=\exp(\rmi\pi)$, $t\coloneqq R^n\rme^{\rmi n (\t+\pi)}$ and in the last step we expanded for large $R$ (\cite{NIST}, formula 8.11.2 with $a=0$ and $u_0=0$ or formul\ae\ 6.11.1 and 6.12.1). 

In the case of the asymptotically polynomial form factors \Eq{H2},
\ban
I_R&=&-\frac{1}{2\pi}\int_{\t^-}^{\t^+}\rmd\t\,\frac{\rme^{-\G[0,(-R\,\rme^{\rmi\t})^n]-\g_{\rm E}}}{(-R\,\rme^{\rmi\t})^n}+O(R^{-1})\nn
&=&-\frac{1}{2\pi R^{n}}\int_{\t^-}^{\t^+}\rmd\t\,\rme^{-\rmi n(\t+\pi)}\rme^{-\G[0,R^n\rme^{\rmi n (\t+\pi)}]-\g_{\rm E}}+O(R^{-1})\nn
&\simeq& -\frac{1}{2\pi R^{n}}\int_{\t^-}^{\t^+}\rmd\t\,\rme^{-\rmi n(\t+\pi)}\exp\left[-\frac{\rme^{-R^n\rme^{\rmi n (\t+\pi)}}}{R^n\rme^{\rmi n (\t+\pi)}}-\g_{\rm E}\right]\label{expexp}\\
&=& \frac{1}{2\pi R^{2n}}\int_{\t^-}^{\t^+}\rmd\t\,\rme^{-2\rmi n(\t+\pi)}\rme^{-R^n\rme^{\rmi n (\t+\pi)}-\g_{\rm E}}+O(R^{-n})\nn
&\simeq& \frac{1}{2\pi R^{2n}}\rme^{-\g_{\rm E}}\int_{\t^-}^{\t^+}\rmd\t\,\rme^{-R^n\rme^{\rmi n (\t+\pi)}}\,,
\ean
where in the third line we expanded the incomplete gamma function inside the integral for large $R$ (see \cite[formula 6.12.1]{NIST} and \cite[chapter 4, section 2]{Olv97}), while in the fourth line we further expanded for large $R$ assuming $|\t+\pi|\leq 3\pi/(2n)$. The condition $|\t+\pi|\leq 3\pi/(2n)$ is abandoned from now on by analytic continuation. The last line yields the same integral as in \Eq{Eiint} up to some prefactors. Note, in one of the intermediate steps, the typical explosive $\exp(\exp)$ behaviour of these form factors outside the conical region.

Therefore, for both classes of form factors we have
\be\label{eR}
I_R\stackrel{R\to+\infty}{\longrightarrow} 0\qquad \Longleftrightarrow\qquad \frac{\rme^{-R^n\rme^{\rmi n (\t^\pm+\pi)}}}{R^{n}}\longrightarrow 0\,.
\ee
Since $n> 0$, the denominator always favours convergence, while the numerator goes to zero or to a constant if, and only if,
\ben
\rme^{-R^n\cos[n (\t^\pm+\pi)]}\longrightarrow {\rm const}\qquad \Longleftrightarrow\qquad \cos[n (\t^\pm+\pi)]\geq 0\,.
\een
This condition holds when\footnote{If, instead of $[-\pi/2,\pi/2]$ (mod $2\pi$), one takes the disjointed intervals $[0,\pi/2]\cup[3\pi/2,2\pi]$ (mod $2\pi$), then one is unable to solve for $\t^\mp$ with $n\in\mathbb{N}$, $m\in\mathbb{Z}$.}
\ben
-\frac{\pi}{2}+2m\pi\leq n (\t^\pm+\pi) \leq \frac{\pi}{2}+2m\pi\,,\qquad m\in\mathbb{Z}\,,
\een
that is,
\ben%\label{condt12}
\left(2m-n-\frac{1}{2}\right)\frac{\pi}{n}\leq \t^\pm\leq \left(2m-n+\frac{1}{2}\right)\frac{\pi}{n}\,,\qquad m\in\mathbb{Z}\,.
\een
Since we are looking for the maximal opening angle of each arc and the latter are counter-clockwise, we identify $n$ wedges $\cC_m$ delimited by
\be\label{ide1}
\boxd{\t^\pm_m\coloneqq (2m-n)\frac{\pi}{n}\pm\Theta\,,\qquad \Theta\coloneqq\frac{\pi}{2n}\,,\qquad m=0,1,\dots,n-1\,,}%\left(2m-n\pm\frac{1}{2}\right)\frac{\pi}{n}
\ee
which enter the definition of the domain of validity of the form factor $\exp[-\H(z)]$ in the complex plane:
\be\label{Ccond}
\boxd{\rme^{-\H(z)}:\qquad \cC\cup\p\cC\,,\qquad\cC=\bigcup_{m=0}^{n-1}\cC_m\,,\qquad \cC_m=\left\{\t^-_m<\arg\,z<\t^+_m\right\}\,.}
\ee
The cone boundary
\be
\p\cC=\left\{z\,\big|\,\arg\,z=\t^\pm_m\right\}
\ee
marks a transition in the behaviour of the numerator in \Eq{eR}. Inside the cone, the numerator is exponentially suppressed, $\rme^{-a R^n}/R^{n}$, where $0<a<1$. On the boundary $\p\cC=\cup_m\p\cC_m$, the numerator is oscillating and bounded and the expression in \Eq{eR} is suppressed at large $R$ as $|\rme^{\rmi(\mp R)^n}/R^{n}|=1/R^{n}$. Outside the cone, the numerator explodes exponentially and one cannot close the contour at infinity.

%%%%%%%%%%%%%%%%%%%%%%%%%%%%%%%%%%%%%%%%%%%%%%%%%%%%%%%%%%%%%%%%%%%%%%%%%%%%%

\subsubsection{Conical region}\label{conic}

From \Eqq{ide1}, we find the opening angle of each conical region $\cC_m$,
\be\label{Thetan}
\t^+_m-\t^-_m=2\Theta=\frac{\pi}{n}\,.
\ee
Note that the demi-cone from $\pi-\Theta$ to $\pi+\Theta$ symmetric with respect to the negative real axis,
\be\label{demi-}
\cC_-=\cC_{0}=\left\{\pi-\frac{\pi}{2n}<\arg\,z<\pi+\frac{\pi}{2n}\right\}\,,
\ee
is always present for any $n$, while the demi-cone from $-\Theta$ to $\Theta$ symmetric with respect to the positive real axis,
\be\label{demi+}
\cC_+=\cC_{\frac{n}{2}}=\left\{-\frac{\pi}{2n}<\arg\,z<\frac{\pi}{2n}\right\}\,,
\ee
is present only when $n$ is even. In particular,
\ba
\hspace{-1.2cm}&&n=1:\qquad \cC=\cC_-=\left\{\frac{\pi}{2}<\arg\,z<\frac{3\pi}{2}\right\},\label{Watcone}\\
\hspace{-1.2cm}&&n=2:\qquad \cC=\cC_+\cup\cC_-=\left\{-\frac{\pi}{4}<\arg\,z<\frac{\pi}{4}\right\}\cup\left\{\frac{3\pi}{4}<\arg\,z<\frac{5\pi}{4}\right\},\label{Krascone}\\
\hspace{-1.2cm}&&n=3:\qquad \cC=\left\{\frac{5\pi}{6}<\arg\,z<\frac{7\pi}{6}\right\}\cup\left\{-\frac{\pi}{2}<\arg\,z<-\frac{\pi}{6}\right\}\cup\left\{\frac{\pi}{6}<\arg\,z<\frac{\pi}{2}\right\},
\ea
and so on. 

For the string-field-theory-like Wataghin form factor (\Eq{H1} with $n=1$) and for Kuz'min form factor (\Eq{H2} with $n=1$), we get $\Theta=\pi/2$ and a conical region corresponding to the whole $\Re\,z<0$ half-plane. In particular, the wedge \Eq{Watcone} is the same found in \cite{Kuzmin:1989sp} (eq.\ (8) and below (10) therein). For Krasnikov form factor (\Eq{H1} with $n=2$) and Tomboulis form factor (\Eq{H2} with $n=2$), we get instead $\Theta=\pi/4$ and the cone $\cC_+\cup\cC_-$. In particular, \Eq{Krascone} agrees with the conical region selected in \cite{Tomboulis:1997gg} (eq.\ (4.2) and below (4.11) therein with the parameter $n$ set to 1). The above derivation is perhaps more transparent than the one in those seminal papers, not only because it makes explicit the direct origin of the conical region with a pedagogical calculation, but also because, in the case of asymptotically polynomial form factors, it unifies into a single expression the somewhat specialized parametrizations used by Kuz'min and Tomboulis for their form factors.

\begin{figure}%[!h]
	\bc
	\includegraphics[width=7cm]{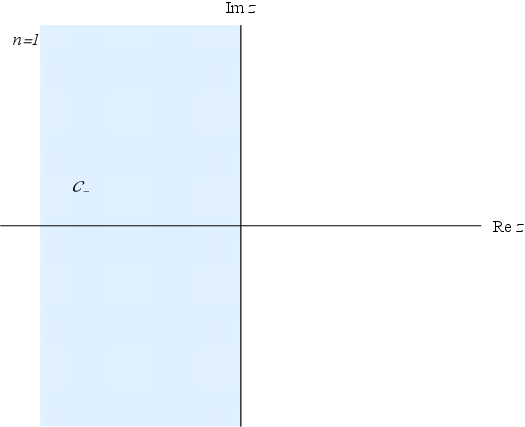}\quad\includegraphics[width=7cm]{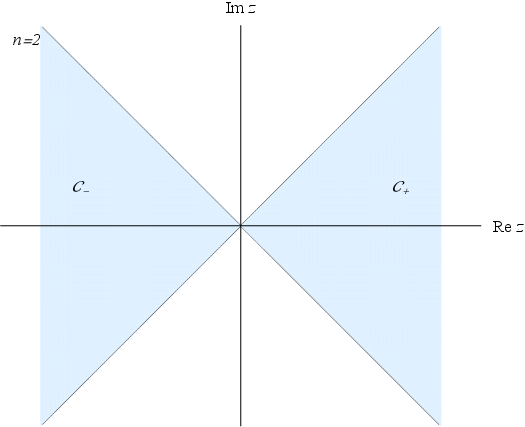}\\\medskip
	\includegraphics[width=7cm]{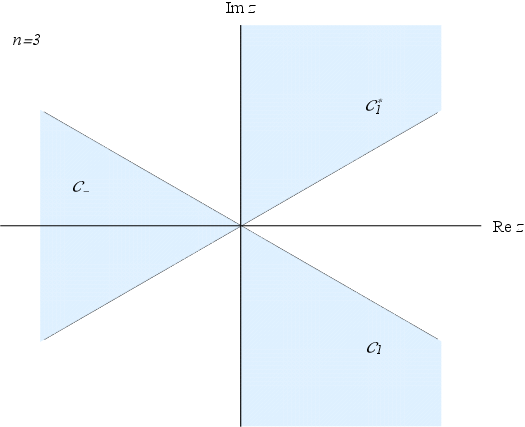}\quad\includegraphics[width=7cm]{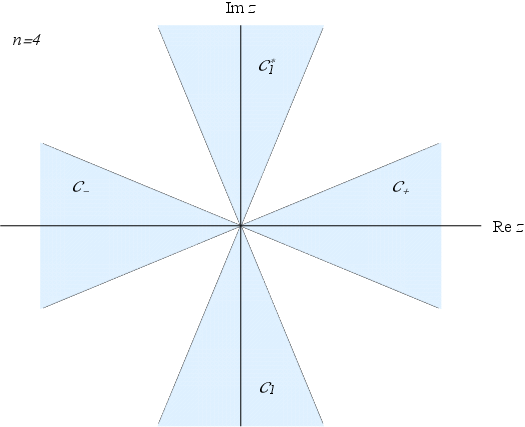}
	\ec
	\caption{\label{fig1} Conical region \Eq{Ccondj} for $n=1,2,3,4$ (shaded area).}
\end{figure}

For general $n$, we can establish a clear pattern of wedges on the complex plane, made of $\cC_\pm$  plus pairs of wedges symmetric with respect to the real axis. These pairs are characterized by complex conjugate phases. While there is no natural $m'$ such that $\t_m^\pm = -\t_{m'}^\pm$ (mod $2\pi$), the condition
\be\label{mmpr}
\t_m^\pm = -\t_{m'}^\mp+2\ell\pi\,,\qquad \ell=0,1\,,\qquad m'=0,1,\dots,n-1\,,
\ee
admits the solution $m'=(1+\ell)n-m$, i.e., $m=0=m'$ for $\ell=-1$ and $m'=n-m$ for $\ell=0$. Therefore, we have two cases depending on $n$.
\begin{itemize}
	\item $n$ odd:
	\ba
	&&\t_0^+=-\t_0^-+2\pi=\pi+\Theta\,,\nn
	&&\t_m^\pm=-\t_{n-m}^\mp\,,\qquad m=1,\dots,n-1\,,\label{tnodd}
	\ea
	so that $m=0$ corresponds to the wedge $\cC_-$ and we can rewrite \Eq{Ccond} as the union of one self-conjugate wedge and $(n-1)/2$ conjugate pairs,
	\be\label{Ccondodd}
	\cC=\cC_-\cup\bigcup_{m=1}^{\frac{n-1}{2}}\left(\cC_m\cup\cC_m^*\right)\,,\qquad \cC_m^*=\cC_{n-m}\,.%\qquad \cC_m=\left\{\t^-_m<\arg\,z<\t^+_m\right\}
	\ee
	\item $n$ even:
	\ba
	&&\t_\frac{n}{2}^+=-\t_\frac{n}{2}^-=\Theta\,,\qquad \t_0^+=-\t_0^-+2\pi=\pi+\Theta\,,\nn
	&&\t_m^\pm=-\t_{n-m}^\mp\,,\qquad \frac{n}{2}\neq m=1,\dots,n-1\,,\label{tneven}
	\ea
	so that $m=0$ and $m=n/2$ correspond, respectively, to the wedges $\cC_-$ and $\cC_+$ and we can rewrite \Eq{Ccond} as the union of two self-conjugate wedges and $n/2-1$ conjugate pairs,
	\be\label{Ccondeven}
	\cC=\cC_-\cup\cC_+\cup\bigcup_{m=1}^{\frac{n}{2}-1}\left(\cC_m\cup\cC_m^*\right)\,.%\qquad \cC_m=\left\{\t^-_m<\arg\,z<\t^+_m\right\}
	\ee
\end{itemize}
We can summarize the decompositions \Eq{Ccondodd} and \Eq{Ccondeven} into the general expression
\be\label{Ccondj}
\boxd{\cC=\cC_-\cup\tilde\cC_+\cup\bigcup_{m=1}^{\lfloor\frac{n-1}{2}\rfloor}\left(\cC_m\cup\cC_m^*\right)\,,}
\ee
where $\lfloor\cdot\rfloor$ is the floor function and we put a tilde on $\cC_+$ to remind ourselves that this wedge is present only when $n$ is even. Figure \ref{fig1} shows the $n=1,2,3,4$ cases.

Notice that this decomposition has a neat correspondence with the oft-used representation of nonlocal operators $f(\B)=\prod_{i=0}^\infty(\B-\om_i^2)$ in terms of quanta of rest mass $\om_i$, some (or most, or all) of which can be purely imaginary \cite{Pais:1950za}. A Cauchy representation reproducing this structure is presented in appendix \ref{appA}. A tentative physical interpretation of the master expression \Eq{Itot} is that it represents the sum of complex conjugate pairs corresponding to virtual particles that never go on-shell.

Two remarks are in order before moving on. First, from \Eq{eR} it is easy to see that the domain of convergence of $\exp(+\H)$ requires $\cos[n(\t^\pm+\pi)]\leq 0$, which results in a conical region complementary to $\cC$, i.e., the white areas in figure~\ref{fig1}:
\be\label{Ccondplus}
\boxd{\rme^{\H(z)}:\qquad \cB=\mathbb{C}\setminus\cC\,,\qquad \tau^\pm_m\coloneqq (2m-n+1)\frac{\pi}{n}\pm\Theta\,,}
\ee
where $\tau^\pm_m=\t^\pm_m+2\Theta$ are the opening angles replacing $\t^\pm_m$. In particular, for $n=1$ one has $\tau^\pm_0=\pm\Theta$ and $\cB$ is the $\Re\,z>0$ half plane; for $n=2$, $\tau^\pm_0=-\pi/4,-3\pi/4$, $\tau^\pm_1=3\pi/4,\pi/4$ and $\cB$ is made of the cones of opening $\pi/2$ along the imaginary axis; and so on.

Second, the domain of convergence of the tree-level propagator and of any Feynman diagram at any perturbative order is the same of the form factor $\exp(-\H)$, or of the form factor $\exp(+\H)$, depending on which operator is regarded as more fundamental. For example, and using a very colloquial terminology, if we take the action as our reference, then $\exp(+\H)$ could be thought of as the main building block, so that the tree-level propagator would scale as $1/\exp(\H)$ (instead of $\exp(-\H)$) and Feynman amplitudes would feature ratios such as $\exp(\H)/\exp(\H)$ instead of $\exp(-\H)/\exp(-\H)$. In any case, the calculation of momentum integrals would be defined on the domain of the main building block, in this example $\cB$. This is only a matter of convention and does not change the physics whatsoever.

%%%%%%%%%%%%%%%%%%%%%%%%%%%%%%%%%%%%%%%%%%%%%%%%%%%%%%%%%%%%%%%%%%%%%%%%%%%%%
%%%%%%%%%%%%%%%%%%%%%%%%%%%%%%%%%%%%%%%%%%%%%%%%%%%%%%%%%%%%%%%%%%%%%%%%%%%%%

\section{Generalized spectral and Källén--Lehmann representations}\label{sec3c}

We are now ready to find the spectral and Källén--Lehmann representations in nonlocal theories with entire form factors and to compare them. 

%%%%%%%%%%%%%%%%%%%%%%%%%%%%%%%%%%%%%%%%%%%%%%%%%%%%%%%%%%%%%%%%%%%%%%%%%%%%%

\subsection{Generalized spectral representation}\label{gespera}

If we tried to apply the Cauchy integral \Eq{opt} on an adaptation of the contour $\G$ 
in the right-hand side of the graphical equation \Eq{equi}, we would immediately face problems. The outer circle of radius $R\to+\infty$ would be replaced by the domain $\cC$ of the form factor $\exp(-\H)$, which is made of $n$ wedges of opening $2\Theta=\pi/n$ as in figure~\ref{fig1}. Then, one would have to include the contribution of the boundary of this conical region, which is described in appendix \ref{appA}. This is not the spectral or the Källén--Lehmann representation of the propagator precisely because the contour cannot be squeezed along the real axis. In fact, this representation does not give transparent information on the physical spectrum.

The alternative is to use the representation \Eq{optm} with
\be\label{KLfinal}
\tilde G(z,k^2)=\frac{\rme^{-\H(-z-k^2)}}{z+k^2}\,.
\ee
The contour is the one depicted in the right-hand side of \Eq{equi2},
\be\label{equi2bis}
\tilde G(-k^2)=\parbox{6.5cm}{\includegraphics[width=6.5cm]{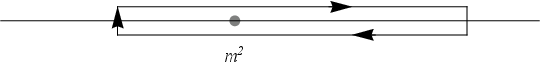}}\,,
\ee
and the calculation is the same as in \Eq{rhs} up to a factor $\exp[-\H(s+k^2)]$ in the integrand. The final result is \Eqq{intfin},
\be\label{kalefinb}
\tilde G(-k^2)=\int_{0}^{+\infty}\rmd s\,\frac{\rme^{-\H(-s-k^2)}}{s+k^2-\rmi\e}\,\rho_{\rm tree}(s)\,,\qquad \rho_{\rm tree}(s)=\de(s-m^2)\,.
\ee
The only subtle point is an extra prescription on the ordering of the limits $\ve\to 0^+$ and $R\to+\infty$ in the analogue of \Eq{IveR}, the contribution of the rightmost short side $I_{\ve,R}$ of the rectangular contour in the right-hand side of \Eq{equi2}. If $n$ is odd, then the contour falls outside the conical region and the length $2\ve$ of $I_{\ve,R}$ must be sent to zero before sending its position $z=R$ to infinity:
\ba
\frac{1}{2\pi\rmi}\int_{I_{\ve,R}}\rmd z\,\frac{\tilde G(z)}{z+k^2}&=& \lim_{R\to+\infty}\lim_{\ve\to 0^+} \int_{-\ve}^{\ve}\rmd t\,\frac{\rme^{-\H(-R-\rmi t-k^2)}}{-R-\rmi t-k^2}\frac{1}{m^2-R-\rmi t}\nn
&\simeq& \lim_{R\to+\infty}\left[\lim_{\ve\to 0^+}\frac{\ve\,\rme^{-\H(-R-k^2)}}{\pi(R+k^2)(m^2-R)}\right]=0\,.\label{IveR2}
\ea
Therefore, strictly speaking, the integral $\int_0^\infty\rmd s$ in \Eq{kalefinb} is calculated as $\lim_{R\to\infty}\int_0^R\rmd s$.

%%%%%%%%%%%%%%%%%%%%%%%%%%%%%%%%%%%%%%%%%%%%%%%%%%%%%%%%%%%%%%%%%%%%%%%%%%%%%

\subsection{Generalized Källén--Lehmann representation}\label{Tprod}

The generalized integral representation \Eq{kalefinb} in the free-field case is not the Källén--Lehmann representation of the nonlocal theory, since, as we will now show, the left-hand side is not the time-ordered (Fourier transform of the) two-point correlation function describing a process from an initial to a final asymptotic state. This correlation function of two fields is defined as
\ba
\Delta_{\rm to}(x-y)&\equiv& \langle\Omega\vert T \phi(x) \phi(y)\vert \Omega \rangle\nn
&\coloneqq& \theta(x^0-y^0)\langle\Omega\vert \phi(x) \phi(y)\vert \Omega \rangle+ \theta(y^0-x^0)\langle\Omega\vert \phi(y) \phi(x)\vert \Omega \rangle\, ,\label{propagator1}
\ea
where ``to'' stands for time ordering, $\vert \Omega \rangle$ is the vacuum state of the nonlocal theory, $T$ is the time ordering operator, $\theta$ is Heaviside step function and $\phi(x)$ is the  nonlocal field operator. Time ordering depends only on the sign of the difference $x^0-y^0$ of the time components of two $D$-dimensional points $x$ and $y$, which is mathematically well-defined even in nonlocal theories experiencing causality violations (see a discussion on the latter in \cite{Alebastrov:1973np,Carone:2016eyp,Boos:2018kir,Boos:2019fbu,Boos:2019zml,Briscese:2019twl}). The problem, however, is to understand how the object \Eq{propagator1} is related to the propagator of the theory appearing in Feynman integrals and scattering amplitudes.

In the standard local free theory, the quantity in (\ref{propagator1}) is equal to the Feynman propagator
\begin{equation}\label{local propagator}
\Delta_{\rm F}^{\rm loc}\left(x-y,m^2\right)\coloneqq \rmi G_{\rm loc}(x-y)\,,\qquad G_{\rm loc}(x)=\int \frac{\rmd^4k}{\left(2\pi\right)^4}\frac{1}{k^2+m^2-\rmi\e} \rme^{-\rmi  k\cdot x} .
\end{equation}
Since the local and nonlocal theories are equivalent if interactions are switched off, as discussed in section \ref{sec3a}, we expect that the two-point function \Eq{propagator1} in the free nonlocal theory will be equal to \Eq{local propagator}, modulo a field redefinition. As we will see, this intuition is right. 

Let us proceed to calculate the two-point correlation function (\ref{propagator1}) in $D=4$ dimensions following the same steps of \cite[section 7.1]{PeSc}. We use the completeness relation to express the identity on the states-space of the theory as
\begin{equation}\label{identity operator}
	\mathbbm{1} = \vert \Omega \rangle \langle \Omega \vert + \sum_\lambda \int \frac{\rmd^3\bm{p}}{\left(2\pi\right)^3} \frac{1}{2 E_p(\lambda)} \vert \lambda_p \rangle  \langle \lambda_p \vert \, ,
\end{equation}
where $ \vert\lambda_p\rangle$ is a state of three-momentum $\bm{p}$ and energy $E_p(\lambda)= \sqrt{\vert \bm{p}\vert^2+m_\lambda^2}$, which is also obtained by a boost of momentum $\bm{p}$ of a zero-momentum state $\vert \lambda_0 \rangle$. Here we used the symbol $p$ for the momenta instead of $k$ to emphasize that we are working on-shell. Just like in standard quantum field theory, the states $|\lambda_p \rangle$ are single-particle, multi-particle and also bound and, in particular, \Eq{identity operator} encodes all possible ways to compose Feynman diagrams at all orders in the perturbative expansion. Thus, at tree level one only has one-particle states $|1_{\rm p}\rangle=|p\rangle$ corresponding to single particles of momentum $p$, plus the states corresponding to interactions. For example, in a $\phi^3$ theory these extra states are $|\lambda_p \rangle=|p\rangle\otimes |p_1,p-p_1\rangle$, where $|p_1,p-p_1\rangle$ is a two-particle state with momenta $p_1$ and $p-p_1$. For the two-point function at the tree level, the only states contributing to the completeness relation are $|1_{\rm p}\rangle$; at one-loop level, also $|p\rangle\otimes |k_1,p-k_1\rangle$ appears in the completeness relation, where $k_1$ and $p-k_1$ are off-shell; and so on.
	
Inserting the identity $\mathbbm{1}$ between the two fields in the correlation function $\langle\Omega\vert \phi(x) \phi(y)\vert \Omega \rangle$ (no time ordering yet), one has 
\begin{equation}\label{propagator2}
	\langle\Omega\vert \phi(x) \phi(y)\vert \Omega \rangle 
	=  \sum_\lambda \int \frac{\rmd^3\bm{p}}{\left(2\pi\right)^3} \frac{1}{2 E_p(\lambda)} \langle\Omega\vert \phi(x)\vert \lambda_p \rangle \langle \lambda_p \vert \phi(y)\vert \Omega \rangle \, ,
\end{equation}
where we have dropped a term $\langle\Omega\vert \phi(x) \vert \Omega \rangle \langle \Omega \vert  \phi(y)\vert \Omega \rangle$ by symmetry. We can express the operator $\phi(x)$ in \Eq{propagator2} in terms of the operator $\tilde\phi(x)$ using the field redefinition \Eq{fire},
\be\label{fire2}
\phi(x) = \rme^{-\frac{1}{2}\H(\Box-m^2)} \, \tilde\phi(x) \,,
\ee
obtaining
\be \label{fire3}
	\langle\Omega\vert \phi(x)\vert \lambda_p \rangle = \langle\Omega\vert \rme^{-\frac{1}{2}\H(\Box-m^2)} \, \rme^{-\rmi \hat p\cdot x} \tilde \phi(0)  \, \rme^{\rmi  \hat p\cdot x}\vert \lambda_p \rangle\, ,
\ee
where $\hat p$ is the momentum operator in the states space, which allows one to write $\tilde\phi(x)= \rme^{-\rmi \hat p\cdot x} \tilde \phi(0)\, \rme^{\rmi  \hat p\cdot x}$. Since $\left[\Box,\hat p\right]=0$, and using the fact that $\vert \lambda_p \rangle$ is an eigenstate of momentum $p$, one has
%\ba
%	\langle\Omega\vert \phi(x)\vert \lambda_p \rangle &=& \langle\Omega\vert \rme^{-\rmi \hat p\cdot x} \rme^{-\frac{1}{2}\H(\Box-m^2)} \,  \tilde \phi(0) \,  \rme^{\rmi   p\cdot x}\vert \lambda_p \rangle\big|_{p^0= E_p(\lambda)}\nn
%	&=&\langle\Omega\vert \rme^{-\rmi \hat p\cdot x} \rme^{-\frac{1}{2}\H(-p^2-m^2)} \,  \tilde \phi(0) \,  \rme^{\rmi   p\cdot x}\vert \lambda_p \rangle\big|_{p^0= E_p(\lambda)}\label{fire4}
%\ea
\be\label{fire4}
\langle\Omega\vert \phi(x)\vert \lambda_p \rangle = \langle\Omega\vert \rme^{-\frac{1}{2}\H(\B-m^2)} \, \rme^{-\rmi \left(\hat p-p\right)\cdot x} \tilde \phi(0)\vert \lambda_p \rangle\big|_{p^0= E_p(\lambda)} %\,,
\ee
(note the replacement of the momentum operator $\hat p$ with the value $p$ of the momentum of the state $\vert \lambda_p \rangle$). Finally, using the Lorentz invariance of  $\vert \Omega \rangle$ and  $\phi(0)$, and acting with the nonlocal operator $\rme^{-\frac{1}{2}\H(\Box-m^2)}$ on the exponential in (\ref{fire4}), keeping in mind that for this state $p^0= E_p(\lambda)$, we have
\ba
\langle\Omega\vert \phi(x)\vert \lambda_p \rangle  &=& \langle\Omega\vert  \rme^{-\frac{1}{2}\H[(\hat p-p)^2-m^2]} \rme^{-\rmi  \left(\hat p-p\right)\cdot x} \tilde \phi(0)  \vert \lambda_p \rangle\big|_{p^0= E_p(\lambda)}\nn
&=& \langle\Omega\vert   \tilde \phi(0)  \vert \lambda_p \rangle \rme^{-\frac{1}{2}\H(m^2_\la-m^2)} \rme^{\rmi   p\cdot x}\big|_{p^0= E_p(\lambda)} \, .\label{fire5}
\ea
Using again the Lorentz invariance of $\vert \Omega \rangle$ and $\phi(0)$, we have
\begin{equation}\label{fire6}
	\langle\Omega\vert \phi(x)\vert \lambda_p \rangle = \langle\Omega\vert   \tilde \phi(0)  \vert \lambda_0 \rangle\rme^{-\frac{1}{2}\H(m^2_\la-m^2)} \rme^{\rmi   p\cdot x}\big|_{p^0= E_p(\lambda)} \, .
\end{equation}
Replacing (\ref{fire6}) into (\ref{propagator2}), we get
\be\label{propagator3}
	\langle\Omega\vert \phi(x) \phi(y)\vert \Omega \rangle 
	=  \sum_\la \vert\langle\Omega\vert   \tilde \phi(0)  \vert \lambda_0 \rangle\vert^2\rme^{-\H(m^2_\la-m^2)}\int \frac{\rmd^3\bm{p}}{\left(2\pi\right)^3} \frac{\rme^{\rmi   p\cdot \left(x-y\right)}}{2 E_p(\lambda)} \Bigg|_{p^0= E_p(\lambda)} \, .
\ee

At this point, one would like to recast this expression as a four-dimensional Lorentzian integral. Since the exponential function in \Eq{propagator3} does not depend on $p$, we can indeed do this in the same way done in the standard local theory, obtaining%. First, we recall that from
\be\label{propagator4}
\Delta_{\rm to}(x-y) 
	=  \sum_\lambda \vert\langle\Omega\vert   \tilde \phi(0)  \vert \lambda_0 \rangle\vert^2\rme^{-\H(m^2_\la-m^2)}\int \frac{\rmd^4k}{\left(2\pi\right)^4} \frac{\rmi}{k^2+m_\lambda^2-\rmi\e}\,  \rme^{\rmi   k \cdot\left(x-y\right)}\,.
\ee
This can be recast in the usual Källén--Lehmann form as
\begin{equation}\label{propagator5}
\Delta_{\rm to}(x-y) 
	=  \int_0^{+\infty} \rmd s \, \Delta_{\rm F}^{\rm loc}\left(x-y,s\right) \,\rho(s) \,,
\end{equation}
where we used \Eq{local propagator} and $\rho(s)$ is the spectral density of the nonlocal theory:
\be\label{rholoc}
\rho(s)	=  \rme^{-\H(s-m^2)}\sum_\lambda \delta(s-m_\lambda^2) \, \vert\langle\Omega\vert   \tilde \phi(0)  \vert \lambda_0 \rangle\vert^2\,,
\ee
which is positive as required by the unitarity of the theory. In momentum space, we obtain \Eq{kalefinint},
\be\label{Gto}
\tilde G_{\rm to}(-k^2)=  \int_0^{+\infty} \rmd s \, \frac{\rho(s)}{s+k^2-\rmi\e}\,,\qquad \rho(s)=\Eq{rholoc}\,,
\ee
where $\tilde G_{\rm to}$ is the Fourier transform of $G_{\rm to}(x-y)=-\rmi\Delta_{\rm to}(x-y)$.
Note that this expression holds also in the presence of interactions.

\subsubsection{Free theory}

Now we verify that the two-point function (\ref{propagator1}) is equal to the standard Feynman propagator if the theory is free. This comes from the fact that the contribution to the mass spectrum of the free nonlocal theory is just that corresponding to one-particle states $\vert 1_{\rm p} \rangle$. Indeed, 
\ba
\rho_{\rm free}(s) &=& \rme^{-\H(s-m^2)}\sum_{1_{\rm p}} \delta(s-m^2) \, \vert\langle\Omega\vert  \tilde \phi(0)  \vert 1_{\rm p} \rangle\vert^2\nn
&=&  \rme^{-\H(s-m^2)}\delta(s-m^2)=\de(s-m^2)\,,\label{rho nonlocal 1}
\ea
which agrees with \Eq{kalefinb} when on-shell. Of course, in ALQG this expression is also equivalent to the local spectral density $\rho=\de(s-m^2)$, since $\H(0)=0$. We have thus recovered \Eq{kalefin} as announced in the introduction. In other NLQGs such as fractional gravity \cite{Calcagni:2022shb}, $\H(0)\neq 0$ and the last step in \Eq{rho nonlocal 1} does not hold.

In the free-field case, the result \Eq{propagator5} can be obtained in a shorter way as follows. In the absence of interactions, the nonlocal field $\phi$ is always on-shell (i.e., $p^2=m^2$) and is equal to the local field $\tilde\phi$, which can be expanded as usual as
\begin{equation}\label{local field}
	\tilde \phi(x)= \int \frac{\rmd^3\bm{p}}{\left(2\pi\right)^3 2 E_p} \left(\tilde a_p \rme^{-\rmi p\cdot x}+\tilde a_p^\dag \rme^{\rmi p\cdot x}\right),
\end{equation}
where the four-momentum $p$ is on-shell. This implies that, for free nonlocal fields,
\be\label{propagator6}
\langle\Omega\vert T \phi(x) \phi(y)\vert \Omega \rangle_{\rm free} 
= \langle\Omega\vert T \tilde \phi(x) \tilde \phi(y)\vert \Omega \rangle_{\rm free}  
= \Delta_{\rm F}^{\rm loc}(x-y,m^2)\,.
\ee
This relation is less general than \Eq{propagator5}--\Eq{rholoc}.

\subsubsection{Time ordering in nonlocal theories}

The free-level expression \Eq{local propagator} is related to the nonlocal diagrammatic propagator by some simple steps:
\ba
\Delta_{\rm free}(x-y)&\coloneqq& \rmi\int \frac{\rmd^4k}{\left(2\pi\right)^4}\frac{\rme^{-\H(-k^2-m^2)}}{k^2+m^2-\rmi\e} \rme^{\rmi  k\cdot (x-y)}\nn
&=& \rme^{-\frac{1}{2}\H(\B_x-m^2)}\rme^{-\frac{1}{2}\H(\B_y-m^2)}\int \frac{\rmd^4k}{\left(2\pi\right)^4}\frac{\rmi}{k^2+m^2-\rmi\e} \rme^{\rmi  k\cdot (x-y)}\nn
&=&\rme^{-\frac{1}{2}\H(\B_x-m^2)}\rme^{-\frac{1}{2}\H(\B_y-m^2)}\Delta_{\rm F}^{\rm loc}\left(x-y,m^2\right)\,.\label{dede}
\ea
The legitimacy of the step between the first and the second line is based on the convergence of such integral, which is assumed by hypothesis in nonlocal theories. In fact, the nonlocal form factor is introduced to improve the convergence of the integrals in scattering amplitudes, which are of a form similar to \Eq{dede}. Notice that this expression immediately yields \Eq{intfin} from \Eq{kalefin}. As we will see below, \Eq{dede} can be extended to the general case with interactions:
\be\label{diagpro}
\Delta(x-y) \coloneqq\rme^{-\frac{1}{2}\H(\B_x-m^2)}\rme^{-\frac{1}{2}\H(\B_y-m^2)}\langle\Omega\vert T \tilde\phi(x)\tilde\phi(y)\vert \Omega \rangle\,,
\ee
where we used the field redefinition \Eq{fire2}. Derivatives of correlation functions are still correlation functions and appear also in standard quantum field theory, as, for instance, in the fermionic sector \cite[section 8.5]{Vel94}.

The technical reason of the difference between the diagrammatic propagator $\Delta$ and its time-ordered part $\Delta_{\rm to}$ is that the form factors $\exp(-\H)$ do not commute with the time ordering $T$, since they do not commute with the step functions in \Eq{propagator1}, as already noted in \cite{Tomboulis:2015gfa}:
\ba
\Delta_{\rm to}(x-y) &=& \langle\Omega\vert T \phi(x) \phi(y)\vert \Omega \rangle\nn
 &=& \langle\Omega\vert T \rme^{-\frac{1}{2}\H(\B_x-m^2)}\tilde\phi(x) \rme^{-\frac{1}{2}\H(\B_y-m^2)}\tilde\phi(y)\vert \Omega \rangle\nn
&\neq& \rme^{-\frac{1}{2}\H(\B_x-m^2)}\rme^{-\frac{1}{2}\H(\B_y-m^2)}\langle\Omega\vert T \tilde\phi(x)\tilde\phi(y)\vert \Omega \rangle\nn
&=&\Delta(x-y)\,.\label{tomb}
\ea
The left-hand side of inequality \Eq{tomb} is nothing but the time-ordered expression \Eq{propagator4} and can be evaluated with the above formalism of state decomposition. Taking the local version of \Eq{propagator3} ($\H=0$) as the starting point, this yields exactly \Eq{propagator3} when $x^0>y^0$:
\ba
&&\rme^{-\frac{1}{2}\H(\B_x-m^2)}\rme^{-\frac{1}{2}\H(\B_y-m^2)}\langle\Omega\vert \tilde\phi(x)\tilde\phi(y)\vert \Omega \rangle\nn 
&&\qquad\qquad =\rme^{-\frac{1}{2}\H(\B_x-m^2)}\rme^{-\frac{1}{2}\H(\B_y-m^2)}\sum_\lambda \int \frac{\rmd^3\bm{p}}{\left(2\pi\right)^3} \frac{\vert\langle\Omega\vert   \tilde \phi(0)  \vert \lambda_0 \rangle\vert^2}{2 E_p(\lambda)}  \rme^{\rmi   p\cdot \left(x-y\right)}\big|_{p^0= E_p(\lambda)}\nn
&&\qquad\qquad =\sum_\lambda \int \frac{\rmd^3\bm{p}}{\left(2\pi\right)^3} \frac{\vert\langle\Omega\vert   \tilde \phi(0)  \vert \lambda_0 \rangle\vert^2}{2 E_p(\lambda)} \rme^{-\H(-p^2-m^2)}\rme^{\rmi   p\cdot \left(x-y\right)}\big|_{p^0= E_p(\lambda)}\nn
&&\qquad\qquad =\sum_\lambda \int\frac{\rmd^3\bm{p}}{\left(2\pi\right)^3} \frac{\vert\langle\Omega\vert   \tilde \phi(0)  \vert \lambda_0 \rangle\vert^2}{2 E_p(\lambda)} \rme^{-\H(m_\la^2-m^2)}\rme^{\rmi p\cdot \left(x-y\right)}\big|_{p^0= E_p(\lambda)},\label{421}
\ea
from which one obtains \Eq{propagator4} after reconstructing the Lorentzian integral via the introduction of the time ordering. On the other hand, the right-hand side of inequality \Eq{tomb} can be calculated taking the local version of \Eq{propagator4} as the starting point, and assuming a discrete spectrum (sum over $\la$ instead of an integral):
\ba
\Delta(x-y) &=& \rme^{-\frac{1}{2}\H(\B_x-m^2)}\rme^{-\frac{1}{2}\H(\B_y-m^2)}\langle\Omega\vert T \tilde\phi(x)\tilde\phi(y)\vert \Omega \rangle\nn
&=&\rme^{-\frac{1}{2}\H(\B_x-m^2)}\rme^{-\frac{1}{2}\H(\B_y-m^2)}\sum_\lambda \vert\langle\Omega\vert   \tilde \phi(0)  \vert \lambda_0 \rangle\vert^2\int \frac{\rmd^4k}{\left(2\pi\right)^4} \frac{\rmi}{k^2+m_\lambda^2-\rmi\e} \rme^{\rmi   k \cdot\left(x-y\right)}\nn
&=&\sum_\lambda \vert\langle\Omega\vert   \tilde \phi(0)  \vert \lambda_0 \rangle\vert^2 \int \frac{\rmd^4k}{\left(2\pi\right)^4} \frac{\rmi\, \rme^{-\H(-k^2-m^2)}}{k^2+m_\lambda^2-\rmi\e}\rme^{\rmi   k \cdot\left(x-y\right)},\label{diagrapro}
\ea
which is the diagrammatic nonlocal propagator \Eq{dialocint} in the presence of interactions. In their absence, it reduces to \Eq{intfin}.

When the action of the form factors on $x$ and $y$ is calculated explicitly on the standard Feynman propagator $\Delta_{\rm F}^{\rm loc}(x-y,m^2)$, \Eq{diagrapro} is very complicated \cite{Tomboulis:2015gfa} and the meaning of time ordering is lost, since infinitely many contact terms are generated in the non-causal part $\tilde G_{\rm nc}$ of the Green's function:\footnote{Comparing our expressions as superpositions of Fock states with those of \cite{Tomboulis:2015gfa}, the standard Feynman propagator $\Delta_{\rm F}^{\rm loc}=\rmi G_{\rm loc}$ \Eq{local propagator} is denoted by $\Delta$ in \cite[eqs.\ (5.1) and (5.5)]{Tomboulis:2015gfa}; our nonlocal diagrammatic propagator $\Delta$, given by the Fourier anti-transform of $\rmi\tilde G$ \Eq{prop}, \Eq{intfin}, \Eq{diagrapro} and the right-hand side of inequality \Eq{tomb} (``$\rme^{-\H}T$'' expressions) is $\tilde\Delta$ in \cite[eqs.\ (5.3) and (5.7)--(5.9)]{Tomboulis:2015gfa}; our time-ordered object $\Delta_{\rm to}$, given by the Fourier anti-transform of $\rmi\tilde G_{\rm to}$ \Eq{kalefinint}, \Eq{propagator4}, \Eq{propagator5}, \Eq{421} and the first line of \Eq{tomb} (``$T\rme^{-\H}$'' expressions) is the nonlocal causal propagator $\tilde\Delta_c$ in \cite[eqs.\ (5.9) and (5.10)]{Tomboulis:2015gfa}; and the Fourier anti-transform of the contact terms $\tilde G_{\rm nc}$ in \Eq{GG} are the infinite sum of terms denoted as $\tilde\Delta_{nc}$ in \cite[eqs.\ (5.9) and (5.12)]{Tomboulis:2015gfa}. Note that $\tilde G_{\rm to}=\tilde G_{\rm loc}$ in the free theory only, eqs.\ \Eq{kalefin} and \Eq{propagator6}.}
\be\label{GG}
\tilde G(-k^2)= \tilde G_{\rm to}(-k^2)+\tilde G_{\rm nc}(k)\,,
\ee
corresponding to $\Delta=\Delta_{\rm to}+\Delta_{\rm nc}$ in position space. %To see this, recall the identity (stemming from \Eq{plemsok}) %[CPNTROLLARE, FORSE CI VA UN SEGNO $-$]
%\be
%\int \rmd^{D-1}\bm{k}\,\frac{\rmi}{k^2+m^2-\rmi\e}\rme^{\rmi k\cdot x}= \int \rmd^{D}k\,\rme^{-\rmi k\cdot x}\left[\theta(k^0)\,\theta(x^0)+\theta(-k^0)\,\theta(-x^0)\right](2\pi)\de(k^2+m^2)\,.
%\ee
At the core of this is the property of nonlocal form factors of smearing distributions, in particular, the step functions. In the last line of \Eq{tomb}, such smearing appears in the term $\rme^{-\frac{1}{2}\H(\B_x-m^2)}[\theta(x^0-y^0)\,\rme^{-\rmi k \cdot\left(x-y\right)}]$ and its counterpart $x\leftrightarrow y$.

A physical interpretation of the difference between $\Delta$ and $\Delta_{\rm to}$ could be the following. On one hand, $\Delta(x-y)$ relates two fields at arbitrary points $x$ and $y$, which can even have separation as small as, or smaller than, the length $\lst$ at which violations of micro-causality happen, where $\lst$ is the fundamental scale implicit in the dimensionless argument $\lst^2\B$ of the form factors.\footnote{Scattering amplitudes in nonlocal quantum field theories satisfy the Bogoliubov causality condition \cite{Alebastrov:1973np}. That means that the quantum theory is safe from causality violations. However, some authors \cite{Carone:2016eyp,Boos:2019fbu,Boos:2019zml} argued about the occurrence of causality violations in the classical theory at the nonlocality scale $\lst$. If so, such violations would occur at a scale where the classical theory should break down, while they do not really happen in the quantum nonlocal theory, as discussed in \cite{Alebastrov:1973np}. Moreover, one may wonder whether, in the classical theory,  micro-instances of causality violation could add up to break causality also at macroscopic scales. This cumulative effect, however, does not happen, for instance, in the form of a Shapiro time delay \cite{Giaccari:2018nzr}. Thus, at a classical level a macroscopic observer would not experience violations of macro-causality even if there were violations of micro-causality; see also the discussion in \cite{Briscese:2019twl}.} On the other hand, the time-ordered part $\Delta_{\rm to}(x-y)$ is the correlation function of nonlocal fields calculated on a complete basis of on-shell states, a notion that implies a measurement at scales $\gg\lst$ for particles such that $\lst m\ll 1$. In fact, re-establishing dimensionful units, we have that $\H(z)$ is a function of $z\coloneqq -\lst^2(k^2+m^2)$. Then, the IR limit $\lst^2 k^2\to 0$ coincides with the on-shell limit $-k^2-m^2\to 0$ when $\lst m\ll 1$, since in both cases one has $z\simeq -\lst^2k^2\to 0$. At scales $\gg\lst$, macro-causality is respected and time ordering acquires the usual interpretation.

%%%%%%%%%%%%%%%%%%%%%%%%%%%%%%%%%%%%%%%%%%%%%%%%%%%%%%%%%%%%%%%%%%%%%%%%%%%%%

\subsection{Local limit}\label{loli}

The local limit of the theory is reached in the IR and corresponds to distances much larger than $\lst$ or energies much smaller than $\Lst\coloneqq \lst^{-1}$. Formally, it would correspond to an infinitely small length scale $\lst$ or an infinitely large energy scale $\Lst\coloneqq \lst^{-1}$. In this limit, the argument of $\H$ tends to zero and, thanks to \Eq{Hir} and \Eq{serie}, $\exp\H\to 1$. Then, since the nonlocal terms in the action are proportional to $\left(\exp\H-1\right)/\B$, the dynamics reduces to the one of Stelle gravity \cite{BasiBeneito:2022wux,Buoninfante:2022ild,Koshelev:2023elc}.

In the limit $\lst\to 0$, the factors $\exp\{-\H[-\lst^2(k^2+m^2)]\}$ and $\exp\{-\H[\lst^2(s-m^2)]\}$ in, respectively, the final expressions \Eq{dialocint} and \Eq{kalefinint} tend to 1, so that the generalized spectral and K\"all\'en--Lehmann representations reduce to the standard K\"all\'en--Lehmann representation of the local theory. This limit can be consistently taken at any of the intermediate steps leading to \Eq{dialocint} and \Eq{kalefinint}, since we are always integrating in that portion of the convergence domain of the nonlocal propagator that overlaps with the convergence domain of the local one. 

This simple but important property is due to the fact, explained at the beginning of section~\ref{gespera}, that what we did was not to modify the prescription on the contour \Eq{equi} (right-hand side) in the local theory but, rather, to choose the contour \Eq{equi2bis}, equivalent to \Eq{equi} only in the local case but valid also in the nonlocal one. The integration contour in \Eq{equi2bis} is exactly the same for both the local and the nonlocal theory, regardless of the value of the length $\lst$. The parallel lines are squeezed along the real axis, the mini-segment $I_{\ve,R}$ at infinity always lies in the convergence domain of the form factor and this contour does not undergo any deformation when taking the local limit.

%%%%%%%%%%%%%%%%%%%%%%%%%%%%%%%%%%%%%%%%%%%%%%%%%%%%%%%%%%%%%%%%%%%%%%%%%%%%%
%%%%%%%%%%%%%%%%%%%%%%%%%%%%%%%%%%%%%%%%%%%%%%%%%%%%%%%%%%%%%%%%%%%%%%%%%%%%%

\section{Discussion}\label{sec4}

To summarize, we studied different representations of the propagator of nonlocal field theories. In particular, we have established most general criteria to calculate the spectral and the Källén--Lehmann representations, which are the ones giving information on the physical spectrum in any theory, local or nonlocal. In our terminology, the spectral representation is for the propagator entering Feynman diagrams, while the Källén--Lehmann representation is for the time-ordered part of the propagator.

We found the generalized spectral representation \Eq{intfin} and the generalized Källén--Lehmann representation \Eq{kalefin} of nonlocal theories with entire form factors, valid at the free level, as well as the spectral representation \Eq{dialocint} and the Källén--Lehmann representation \Eq{kalefinint} in the presence of interactions. Since \Eq{intfin} is not in the form \Eq{kale}, Weinberg's theorem \cite{Wei95} does not apply and we reconcile the facts that these theories are free-level unitary and that the UV diagrammatic propagator scales faster than $\sim k^{-2}$. Surprisingly, we have obtained an example of a theory where Weinberg's theorem holds for the Källén--Lehmann representation (since $\Delta_{\rm F}^{\rm loc}\sim k^{-2}$ in momentum space) and, yet, momentum integrals do show a better convergence in the UV than standard two-derivative field theory. The magic is operated by the inequivalence \Eq{GG} between the propagator appearing in Feynman diagrams and the time-ordered two-point correlation function.

From these results, we can understand at two levels of reasoning why the standard representation \Eq{kale} does not hold in nonlocal theories. 

At a mathematical level, when the spectral representation is derived directly from Cauchy's integral representation for the propagator, one must take into account that nonlocal form factors produce a number of terms which mix the $s$ and $k$ dependence in \Eq{intfin} in such a way that cannot be factorized simply as in \Eq{kale}. This technical explanation was already pointed out in \cite{Calcagni:2022shb} for fractional quantum gravity and, as we have seen, holds also for entire form factors in nonlocal field theories and, in particular, ALQG.

A physical understanding comes when noting that the standard Källén--Lehmann representation is a very general consequence of having a time-ordered two-point function \cite[section 24.2.1]{Sch14}. This is the one and only point of the proof of \cite{Sch14} where NLQG, and nonlocal theories in general, differ with respect to local quantum field theory, as we showed in section \ref{Tprod}. It is also responsible of the inequivalence \Eq{tomb}. The step from the on-shell non-time-ordered expression \Eq{propagator3} and the off-shell time-ordered Lorentz-invariant expression \Eq{propagator4} cannot recover the structure of the nonlocal diagrammatic propagator \Eq{prop} because the contour in the complex $k^0$-plane making this step possible does not respect the conical domain region $\cC$ where the Cauchy representation of the form factor is well-defined.

All of this is in pleasant agreement with the field-redefinition argument of section \ref{sec3a} but also goes one step beyond in showing that the generalization to higher loops is just a matter of calculations. The integrand in the spectral and the Källén--Lehmann representations will be more and more complicated at higher orders in perturbation theory but no deviations from the conceptual and mathematical frames set in the present work are expected. Although a painstaking check of perturbative unitarity could be done with this method order by order, it would only confirm what we already know from the general and more elegant proof via the Cutkosky rules \cite{Briscese:2018oyx,Pius:2016jsl}.
	
\medskip

\emph{Note added.} After the completion of this paper, we became aware of \cite{Paganis:2024end}, where the generalized spectral representation \Eq{intfin} for an exponential form factor $\exp\B$ was guessed from asymptotic considerations related to IR LHC phenomenology. Our proof for \emph{any} entire form factor $\exp\H(\B)$ corroborates those findings and also puts on a firmer ground the study of scattering amplitudes in nonlocal field theories, which so far have assumed the profile of the spectral density (for an exponential form factor $\exp\B^n$) without derivation \cite{Efimov:2001ad,Tokuda:2019nqb,Buoninfante:2023dyd}.

%%%%%%%%%%%%%%%%%%%%%%%%%%%%%%%%%%%%%%%%%%%%%%%%%%%%%%%%%%%%%%%%%%%%%%%%%%%%%
%%%%%%%%%%%%%%%%%%%%%%%%%%%%%%%%%%%%%%%%%%%%%%%%%%%%%%%%%%%%%%%%%%%%%%%%%%%%%

\section*{Acknowledgments}

G.C.\ thanks E.\ Pajer for triggering this project with his questions. G.C.\ and L.M.\ are supported by grant PID2020-118159GB-C41 funded by MCIN/AEI/10.13039/501100011033. %L.M.\ is also supported by the Basic Research Program of the Science, Technology, and Innovation Commission of Shenzhen Municipality (grant no.\ JCYJ2018030\-2174206969).

%%%%%%%%%%%%%%%%%%%%%%%%%%%%%%%%%%%%%%%%%%%%%%%%%%%%%%%%%%%%%%%%%%%%%%%%%%%%%
%%%%%%%%%%%%%%%%%%%%%%%%%%%%%%%%%%%%%%%%%%%%%%%%%%%%%%%%%%%%%%%%%%%%%%%%%%%%%

\appendix

\section{Cauchy representation of the form factors}\label{appA}

In this appendix, we compute Cauchy's integral representation \Eq{fofamain} for exponential and asymptotically polynomial form factors with zero mass, $m^2=0$, as well as for the tree-level propagator. We consider the contour enclosing the maximal area in the complex plane,
\be\label{Gammacon}
\G'=\bigcup_{m=0}^{n-1}\left(\G_R^m\cup\p\cC_m\right),\qquad m=0,1\dots,n-1\,,
\ee
where $\p\cC_m$ is the boundary of the conical region $\cC_m$ in which the contribution of the arc $\G_R^m$ at infinity vanishes as explained in section \ref{sec:arcsinf}. All the wedges can be connected at finite $|z|$ to make a single contour on the complex plane but these contributions are identically zero. %The calculation below is similar to that developed for fractional quantum gravity \cite{Calcagni:2022shb} and, in fact, it is simpler, since we do not have branch cuts in nonlocal theories with entire form factors.

Since we do not have any discontinuity at the boundary $\p\cC$ for $z<\infty$ and we integrate radially from zero to infinity, we can integrate exactly on top of such boundary. The integral $I_{\p\cC}$ we are seeking to compute is the sum of the contribution of each wedge, whose boundary is parametrized by
\be
\p\cC_m=\cC_m^-\cup\cC_m^+\,,\qquad \cC_m^\pm\ni z_m^\pm = s\,\rme^{\rmi\t_m^\pm}\,,\qquad s\in[\ve,+\infty)\,,
\ee
where, according to \Eqqs{tneven} and \Eq{tnodd},
\be\label{zz*}
z_0^+=z_0^{-*}\,,\qquad z_{n-m}^\pm=z_m^{\mp*}\,,\qquad m=1,\dots,n-1\,.
\ee

Although the contribution of each wedge is not real-valued except for $\cC_+$ and $\cC_-$ (\Eqqs{demi+} and \Eq{demi-}), their total sum is real, since all boundary angles corresponding to wedges where the integral is not real-valued are paired into complex conjugate phases according to \Eq{mmpr}. Given the structure of the form factor, this is a sufficient condition to ensure reality of $I_{\p\cC}$ \cite{Calcagni:2022shb}. Therefore, using the general decomposition \Eq{Ccondj}, we have
%\ba
%I_{\p\cC} &=& \frac{1}{2\pi\rmi}\sum_{m=0}^{n-1}\int_{\p\cC_m}\rmd z\,\frac{\tilde G(z)}{z+k^2}\nn
%&=& \frac{1}{2\pi\rmi}\sum_{m=0}^{n-1}\left[\int_{\p\cC_m^+}\rmd z_m^+\,\frac{\tilde G(z_m^+)}{z_m^++k^2}+\int_{\p\cC_m^-}\rmd z_m^-\,\frac{\tilde G(z_m^-)}{z_m^-+k^2}\right]\nn
%&=&\lim_{\ve\to 0^+}\sum_{m=0}^{n-1} \int_{\ve}^{+\infty}\rmd s\left[\frac{\rho^+_m(s)}{s+\rme^{-\rmi\t_m^+}k^2}+\frac{\rho^-_m(s)}{s+\rme^{-\rmi\t_m^-}k^2}\right]\,,\label{Gcutgen}
%\ea
%where
%\be
%\rho_m^\pm(s)\coloneqq \mp\frac{1}{2\pi\rmi}\,\tilde G(z_m^\pm)=\pm\frac{1}{2\pi\rmi}\,\frac{\rme^{-\H(s\,\rme^{\rmi\t_m^\pm})}}{s\,\rme^{\rmi\t_m^\pm}}\,.\label{rhogen}
%\ee
\be\label{intot}
I_{\p\cC} =  I_{\p\cC_-}+\frac{1+(-1)^n}{2} I_{\p\cC_+}+\sum_{m=1}^{\lfloor\frac{n-1}{2}\rfloor}I_{\p\cC_m\cup\p\cC_m^*}\,,
\ee

%%%%%%%%%%%%%%%%%%%%%%%%%%%%%%%%%%%%%%%%%%%%%%%%%%%%%%%%%%%%%%%%%%%%%%%%%%%%%

\subsection{Form factor}

The first term is always present and reads
\ba
I_{\p\cC_-}&=& \frac{1}{2\pi\rmi}\left[\int\rmd z_{0}^-\,\frac{\rme^{-\H(z_{0}^-)}}{z_{0}^-+k^2}+\int\rmd z_{0}^+\,\frac{\rme^{-\H(z_{0}^+)}}{z_{0}^++k^2}\right]\nn
&\stackrel{\textrm{\tiny\Eq{tnodd}}}{=}& \lim_{\ve\to 0^+}\frac{1}{2\pi\rmi}\int_{\ve}^{+\infty}\rmd s \left\{\frac{\rme^{-\H[s\,\rme^{\rmi(\pi-\Theta)}]}}{s+\rme^{-\rmi(\pi-\Theta)}k^2}-\frac{\rme^{-\H[s\,\rme^{\rmi(\pi+\Theta)}]}}{s+\rme^{-\rmi(\pi+\Theta)}k^2}\right\}\nn
&=& \lim_{\ve\to 0^+}\frac{1}{2\pi\rmi}\int_{\ve}^{+\infty}\rmd s \left\{\frac{\rme^{-\H[s\,\rme^{\rmi(\pi-\Theta)}]}}{s+\rme^{-\rmi(\pi-\Theta)}k^2}-\frac{\rme^{-\H[s\,\rme^{-\rmi(\pi-\Theta)}]}}{s+\rme^{\rmi(\pi-\Theta)}k^2}\right\}\nn
&=&\int_{0}^{+\infty}\rmd s\,\frac{\rho(s,\pi-\Theta)\,\rme^{\rmi\Psi(s,\pi-\Theta)}\left[s+\rme^{\rmi(\pi-\Theta)}k^2\right]}{\left[s+\rme^{-\rmi(\pi-\Theta)}k^2\right]\left[s+\rme^{\rmi(\pi-\Theta)}k^2\right]}\nn
&&+\int_{0}^{+\infty}\rmd s\,\frac{\rho(s,\pi-\Theta)\,\rme^{-\rmi\Psi(s,\pi-\Theta)}\left[s+\rme^{-\rmi(\pi-\Theta)}k^2\right]}{\left[s+\rme^{-\rmi(\pi-\Theta)}k^2\right]\left[s+\rme^{\rmi(\pi-\Theta)}k^2\right]}\nn
&=& 2\int_{0}^{+\infty}\rmd s\, \rho(s,\pi-\Theta)\,\frac{s\,\cos\Psi(s,\pi-\Theta)+k^2\cos[\Psi(s,\pi-\Theta)+\pi-\Theta]}{s^2+2sk^2\cos(\pi-\Theta)+k^4}\nn
&=&I(\pi-\Theta)\,,\label{conint-}
\ea
where we set the lower extremum to $\ve=0$ because the integrand vanishes in the limit $s\to 0$ and
\ba
&&\rho(s,\vp)\,\rme^{\pm\rmi\Psi(s,\vp)}\coloneqq \pm\frac{1}{2\pi\rmi}\,\rme^{-\H(s\,\rme^{\pm\rmi\vp})}\,,\label{rhogen}\\
&&I(\vp)\coloneqq 2\int_{0}^{+\infty}\rmd s\, \rho(s,\vp)\,\frac{s\,\cos\Psi(s,\vp)+k^2\cos[\Psi(s,\vp)+\vp]}{s^2+2sk^2\cos\vp+k^4}\,.\label{Igen}
\ea

The second term, present only when $n$ is even, is
\ba
I_{\p\cC_+} &=& \frac{1}{2\pi\rmi}\int_{\p\cC_+}\rmd z\,\frac{\rme^{-\H(z)}}{z+k^2}\nn
&=& \frac{1}{2\pi\rmi}\left[\int\rmd z_{n/2}^-\,\frac{\rme^{-\H(z_{n/2}^-)}}{z_{n/2}^-+k^2}+\int\rmd z_{n/2}^+\,\frac{\rme^{-\H(z_{n/2}^+)}}{z_{n/2}^++k^2}\right]\nn
&\stackrel{\textrm{\tiny\Eq{tneven}}}{=}& \frac{1}{2\pi\rmi}\int_{0}^{+\infty}\rmd s \left[\frac{\rme^{-\H(s\,\rme^{-\rmi\Theta})}}{s+\rme^{\rmi\Theta}k^2}-\frac{\rme^{-\H(s\,\rme^{\rmi\Theta})}}{s+\rme^{-\rmi\Theta}k^2}\right]\nn
&=&-\int_{0}^{+\infty}\rmd s\,\frac{\rho(s,\Theta)\,\rme^{-\rmi\Psi(s,\Theta)}\left(s+\rme^{-\rmi\Theta}k^2\right)+\rho(s,\Theta)\,\rme^{\rmi\Psi(s,\Theta)}\left(s+\rme^{\rmi\Theta}k^2\right)}{\left(s+\rme^{\rmi\Theta}k^2\right)\left(s+\rme^{-\rmi\Theta}k^2\right)}\nn
&=& -2\int_{0}^{+\infty}\rmd s\, \rho(s,\Theta)\,\frac{s\,\cos\Psi(s,\Theta)+k^2\cos[\Psi(s,\Theta)+\Theta]}{s^2+2sk^2\cos\Theta+k^4}\nn
&=&-I(\Theta)\,.\label{conint+}
\ea
One can check this also noting that $z_0^\pm=-z_{n/2}^\pm=-z_{n/2}^{\mp*}$. 

The last contribution of \Eq{intot}, present only for $n\geq 3$, is the sum of conjugate wedges. For the $m$-th pair,
\ba
I_{\p\cC_m\cup\p\cC_m^*} &=& \frac{1}{2\pi\rmi}\int_{\p\cC_m}\rmd z\,\frac{\rme^{-\H(z)}}{z+k^2}+ \frac{1}{2\pi\rmi}\int_{\p\cC_m^*}\rmd z\,\frac{\rme^{-\H(z)}}{z+k^2}\nn
&=&\frac{1}{2\pi\rmi}\left[\int\rmd z_{m}^-\,\frac{\rme^{-\H(z_{m}^-)}}{z_{m}^-+k^2}+\int\rmd z_{m}^+\,\frac{\rme^{-\H(z_{m}^+)}}{z_{m}^++k^2}\right]\nn
&&+\frac{1}{2\pi\rmi}\left[\int\rmd z_{n-m}^-\,\frac{\rme^{-\H(z_{n-m}^-)}}{z_{n-m}^-+k^2}+\int\rmd z_{n-m}^+\,\frac{\rme^{-\H(z_{n-m}^+)}}{z_{n-m}^++k^2}\right]\nn
&=&\frac{1}{2\pi\rmi}\left[\int\rmd z_{m}^-\,\frac{\rme^{-\H(z_{m}^-)}}{z_{m}^-+k^2}+\int\rmd z_{m}^+\,\frac{\rme^{-\H(z_{m}^+)}}{z_{m}^++k^2}\right]\nn
&&+\frac{1}{2\pi\rmi}\left[\int\rmd z_{m}^{+*}\,\frac{\rme^{-\H(z_{m}^{+*})}}{z_{m}^{+*}+k^2}+\int\rmd z_{m}^{-*}\,\frac{\rme^{-\H(z_{m}^{-*})}}{z_{m}^{-*}+k^2}\right]\nn
&=& \frac{1}{2\pi\rmi}\int_{0}^{+\infty}\rmd s \left[\frac{\rme^{-\H(s\,\rme^{\rmi\t_m^-})}}{s+\rme^{-\rmi\t_m^-}k^2}-\frac{\rme^{-\H(s\,\rme^{\rmi\t_m^+})}}{s+\rme^{-\rmi\t_m^+}k^2}\right]\nn
&&+ \frac{1}{2\pi\rmi}\int_{0}^{+\infty}\rmd s \left[\frac{\rme^{-\H(s\,\rme^{-\rmi\t_m^+})}}{s+\rme^{\rmi\t_m^+}k^2}-\frac{\rme^{-\H(s\,\rme^{-\rmi\t_m^-})}}{s+\rme^{\rmi\t_m^-}k^2}\right]\nn
&=& \frac{1}{2\pi\rmi}\int_{0}^{+\infty}\rmd s \left[\frac{\rme^{-\H(s\,\rme^{\rmi\t_m^-})}}{s+\rme^{-\rmi\t_m^-}k^2}-\frac{\rme^{-\H(s\,\rme^{-\rmi\t_m^-})}}{s+\rme^{\rmi\t_m^-}k^2}
\right]\nn
&&+\frac{1}{2\pi\rmi}\int_{0}^{+\infty}\rmd s \left[\frac{\rme^{-\H(s\,\rme^{-\rmi\t_m^+})}}{s+\rme^{\rmi\t_m^+}k^2}
-\frac{\rme^{-\H(s\,\rme^{\rmi\t_m^+})}}{s+\rme^{-\rmi\t_m^+}k^2}\right]\nn
&=&\int_{0}^{+\infty}\rmd s\,\frac{\rho(s,\t_m^-)\,\rme^{\rmi\Psi(s,\t_m^-)}\left(s+\rme^{\rmi\t_m^-}k^2\right)+\rho(s,\t_m^-)\,\rme^{-\rmi\Psi(s,\t_m^-)}\left(s+\rme^{-\rmi\t_m^-}k^2\right)}{\left(s+\rme^{-\rmi\t_m^-}k^2\right)\left(s+\rme^{\rmi\t_m^-}k^2\right)}\nn
&&-\int_{0}^{+\infty}\rmd s\,\frac{\rho(s,\t_m^+)\,\rme^{\rmi\Psi_m^+(s)}\left(s+\rme^{-\rmi\t_m^+}k^2\right)+\rho(s,\t_m^+)\,\rme^{-\rmi\Psi_m^+(s)}\left(s+\rme^{\rmi\t_m^+}k^2\right)}{\left(s+\rme^{\rmi\t_m^+}k^2\right)\left(s+\rme^{-\rmi\t_m^+}k^2\right)}\nn
&=& 2\int_{0}^{+\infty}\rmd s\, \rho(s,\t_m^-)\,\frac{s\,\cos\Psi(s,\t_m^-)+k^2\cos[\Psi(s,\t_m^-)+\t_m^-]}{s^2+2sk^2\cos\t_m^-+k^4}\nn
&&-2\int_{0}^{+\infty}\rmd s\,\rho(s,\t_m^+)\, \frac{s\,\cos\Psi(s,\t_m^+)+k^2\cos[\Psi(s,\t_m^+)+\t_m^+]}{s^2+2sk^2\cos\t_m^++k^4}\nn
&=& I(\t_m^-)-I(\t_m^+)\nn%=I(\t_m^-+2\Theta)-I(\t_m^-)\nn
&=& I\left[(2m-n)\frac{\pi}{n}-\Theta\right]-I\left[(2m-n)\frac{\pi}{n}+\Theta\right]\,,\label{conintm}
\ea
where we used the definition \Eq{ide1} of $\t_m^\pm$.

Combining \Eq{intot}, \Eq{conint-} and \Eq{conint+}, the Cauchy representation of the form factor taking as contour $\G$ the maximal one in the complex plane reads
\ba
\rme^{-\H(k^2)}&=& I(\pi-\Theta)-\frac{1+(-1)^n}{2} I(\Theta)\nn
&&+\sum_{m=1}^{\lfloor\frac{n-1}{2}\rfloor}\left\{I\left[(2m-n)\frac{\pi}{n}-\Theta\right]-I\left[(2m-n)\frac{\pi}{n}+\Theta\right]\right\}.\label{Itot}
\ea
From this, we have that
\ba
\hspace{-1cm}\vp = \pi-\Theta,\,\Theta,\, (2m-n)\frac{\pi}{n}\pm\Theta\quad&\Longrightarrow&\quad n\vp=n\pi-\frac{\pi}{2},\,\frac{\pi}{2},\,(2m-n)\pi\widetilde\pm\frac{\pi}{2}\nn
\hspace{-1cm}&\Longrightarrow&\quad \rme^{\pm\rmi n(\pi+\vp)}=\mp\rmi,\,\pm \rmi,\,\pm\widetilde\pm\rmi\,,\label{A11}
\ea
where for $\vp=\Theta$ we used the fact that $n$ is even. Therefore, for the exponential form factors \Eq{H1}, it is not difficult to see that
\ben
\rho(s,\vp)\,\rme^{\pm\rmi\Psi(s,\vp)}=\pm\frac{1}{2\pi\rmi}\,\rme^{-(-s\rme^{\pm\rmi n\vp})^n}=\frac{1}{2\pi}\,\rme^{\mp\rmi\frac{\pi}{2}-s^n\rme^{\pm\rmi n(\pi+\vp)}}\,,
\een
and from \Eq{A11}
\ba
\hspace{-1.5cm}&& \rho(s,\vp)= \frac{1}{2\pi}\,,\\
\hspace{-1.5cm}&&\Psi(s,\pi-\Theta) = -\frac{\pi}{2}+s^n\,,\qquad\Psi(s,\Theta) = -\frac{\pi}{2}-s^n\,,\qquad \Psi(s,\t_m^\pm) = -\frac{\pi}{2}\mp s^n\,.
\ea
 
For the asymptotically polynomial form factors \Eq{H2}, we get
\ben
\rho(s,\vp)\,\rme^{\pm\rmi\Psi(s,\vp)} = \pm\frac{\rme^{-\g_{\rm E}}}{2\pi\rmi}\,\frac{\rme^{-\G[0,(-s\,\rme^{\pm\rmi\vp})^n]}}{(-s\,\rme^{\pm\rmi\vp})^n}=\pm\frac{\rme^{-\g_{\rm E}}}{2\pi\rmi}\,\rme^{\mp\rmi n(\pi+\vp)}\frac{\rme^{-\G[0,s^n\rme^{\pm\rmi (\pi+\vp)}]}}{s^n}\,,
\een
hence
\ba
\hspace{-1.5cm}&\rho(s,\pi-\Theta) =\dfrac{\rme^{-\g_{\rm E}}}{2\pi}\dfrac{\rme^{-\Re\,\G(0,\rmi s^n)}}{s^n}\,,\qquad &\Psi(s,\pi-\Theta) = \Im\,\G(0,\rmi s^n)\,,\label{psiap1}\\
\hspace{-1.5cm}&\rho(s,\Theta) =-\rho(s,\pi-\Theta)\,,\qquad\qquad\hspace{.6cm} &\Psi(s,\Theta) = -\Psi(s,\pi-\Theta)\,,\label{psiap2}\\
\hspace{-1.5cm}&\rho(s,\t_m^\pm) =\mp\rho(s,\pi-\Theta)\,,\qquad\qquad\hspace{.6cm} &\Psi(s,\t_m^\pm) = \mp\Psi(s,\pi-\Theta)\,,\label{psiap3}
\ea
where we used the properties $\Re\,\G(0,-\rmi x)=\Re\,\G(0,\rmi x)$ and $\Im\,\G(0,-\rmi x)=-\Im\,\G(0,\rmi x)$. Note that the phases $\Psi$ are bounded by the largest of the solutions of $\p_x\Im\,\G(0,\rmi x)=\sin x/x=0$, hence $x=s^n=\pi,2\pi$:
\be
-\frac{\pi}{21}\approx-0.1526\approx\Im\,\G(0,2\rmi\pi)\leq\Psi(s,\pi-\Theta)\leq \Im\,\G(0,\rmi\pi)\approx 0.2811\approx \frac{\pi}{11}\,.
\ee
%Curiously, while for exponential form factors the signs of $\cos\Psi$ and $\cos(\Psi+\vp)$ are not positive semi-definite because $\Psi\in[0,2\pi]$, for asymptotically polynomial form factors $\cos\Psi>0$ for any $s$, while $\cos(\Psi+\vp)>0$ if $n\geq 2$.

%%%%%%%%%%%%%%%%%%%%%%%%%%%%%%%%%%%%%%%%%%%%%%%%%%%%%%%%%%%%%%%%%%%%%%%%%%%%%

\subsection{Propagator}

Expression \Eq{Itot} also serves as the Cauchy representation for the propagator upon adding the contribution from the pole in $z=0$,
\be
I_{\rm p}(\Theta) = \frac{1}{2}\frac{1}{k^2}=\frac{1}{\pi}\int_0^{+\infty}\frac{\rmd s}{s^2+k^4}=\frac{n\sin\Theta}{\pi}\int_0^{+\infty}\frac{\rmd s}{s^2+2sk^2\cos\Theta+k^4}\,,\label{poletot}
\ee
which is half the value of the residue, and modifying the definitions of $\rho$ and $\Psi$:
\be
\rho(s,\vp)\,\rme^{\pm\rmi\Psi(s,\vp)}= \pm\frac{1}{2\pi\rmi}\,\frac{\rme^{-\H(s\,\rme^{\pm\rmi\vp})}}{-s\,\rme^{\pm\rmi\vp}}\,.
\ee
For example, for exponential form factors,
\ben
\rho(s,\vp)\,\rme^{\pm\rmi\Psi(s,\vp)}=\frac{1}{2\pi}\frac{1}{s}\,\rme^{\pm\rmi\frac{\pi}{2}\mp\rmi\vp-(-s)^n\rme^{\pm\rmi n\vp}}\,,
\een
so that from \Eq{A11}
\ba
&&\rho(s,\vp)=\rho(s)= \frac{1}{2\pi s}\,,\\
&&\Psi(s,\pi-\Theta) = \Theta-\frac{\pi}{2}+s^n\,,\qquad\Psi(s,\Theta) =-\Psi(s,\pi-\Theta)\,,\qquad
\Psi(s,\t_m^\pm) = \frac{\pi}{2}-\t_m^\pm\mp s^n\,.\nn\label{psiexp}
\ea

However, it is important to stress that, since we did not use the contour wrapped around the real axis as in section \ref{sec4}, this would not be the spectral representation of the propagator and one would be unable to study unitarity from the sign of the integrand in \Eq{Igen}, which is clearly not positive semi-definite since it oscillates between positive and negative values along $s$, due to the range $\Psi\in[0,2\pi]$ taken by the phases in the case \Eq{psiexp}.

%%%%%%%%%%%%%%%%%%%%%%%%%%%%%%%%%%%%%%%%%%%%%%%%%%%%%%%%%%%%%%%%%%%%%%%%%%%%%
%%%%%%%%%%%%%%%%%%%%%%%%%%%%%%%%%%%%%%%%%%%%%%%%%%%%%%%%%%%%%%%%%%%%%%%%%%%%%

\end{document}